\documentclass[]{fairmeta}
\usepackage[]{todonotes}
\usepackage{lato}
\usepackage[sort&compress,numbers,super]{natbib}
\usepackage{url}
\usepackage{macros}
\usepackage{booktabs}
\usepackage{graphicx}
\usepackage{multirow}
\makeglossaries

\newcommand{\cmu}{Department of Chemical Engineering, Carnegie Mellon University}
\newcommand{\first}{These authors contributed equally to this work.}
\newcommand{\utmech}{Department of Mechanical and Industrial Engineering, University of Toronto}
\newcommand{\utel}{Department of Electrical and Computer Engineering, University of Toronto}
\newcommand{\meta}{Fundamental AI Research (FAIR) Meta}
\newcommand{\vsp}{VSParticle BV}
\newcommand{\utmat}{Department of Material Science and Engineering, University of Toronto}
\newcommand{\atmd}{Alliance for AI-Accelerated Materials Discovery (A3MD)}

\title{Open Catalyst Experiments 2024 (OCx24): Bridging Experiments and Computational Models}

\author[1*]{Jehad Abed}
\author[2,5,6*]{Jiheon Kim}
\author[1*]{Muhammed Shuaibi}
\author[3*]{Brook Wander}
\author[7]{Boris Duijf}
\author[4,6]{Suhas Mahesh}
\author[5,6]{Hyeonseok Lee}
\author[1]{Vahe Gharakhanyan}
\author[2,6]{Sjoerd Hoogland}
\author[7]{Erdem Irtem}
\author[1]{Janice Lan}
\author[7]{Niels Schouten}
\author[7]{Anagha Usha Vijayakumar}
\author[4,6]{Jason Hattrick-Simpers}
\author[3]{John R. Kitchin}
\author[1]{Zachary W. Ulissi}
\author[7]{Aaike van Vugt}
\author[2,6\dagger]{Edward H. Sargent}
\author[5,6\dagger]{David Sinton}
\author[1\dagger]{C. Lawrence Zitnick}

\affiliation[1]{\meta} 
\affiliation[2]{\utel}
\affiliation[3]{\cmu}
\affiliation[4]{\utmat}
\affiliation[5]{\utmech}
\affiliation[6]{\atmd}
\affiliation[7]{\vsp}

\abstract{%\lz{Any updates to authors and order, email Larry - zitnick@meta.com}

The search for low-cost, durable, and effective catalysts is essential for green hydrogen production and \cd~upcycling to help in the mitigation of climate change. Discovery of new catalysts is currently limited by the gap between what AI-accelerated computational models predict and what experimental studies produce. To make progress, large and diverse experimental datasets are needed that are reproducible and tested at industrially-relevant conditions. We address these needs by utilizing a comprehensive high-throughput characterization and experimental pipeline to create the Open Catalyst Experiments 2024 (\ocx) dataset. The dataset contains 572 samples synthesized using both wet and dry methods with X-ray fluorescence and X-ray diffraction characterization. We prepared 441 gas diffusion electrodes, including replicates, and evaluated them using zero-gap electrolysis for \cd~reduction (\cdrr) and hydrogen evolution reactions (\her) at current densities up to $300$ mA/cm$^2$. To find correlations with experimental outcomes and to perform computational screens, DFT-verified adsorption energies for six adsorbates were calculated on $\sim$20,000 inorganic materials requiring 685 million AI-accelerated relaxations. Remarkably from this large set of materials, a data driven Sabatier volcano independently identified Pt as being a top candidate for \her~without having any experimental measurements on Pt or Pt-alloy samples. We anticipate the availability of experimental data generated specifically for AI training, such as \ocx, will significantly improve the utility of computational models in selecting materials for experimental screening. }

\date{November 18, 2024}

\correspondence{E. H. Sargent (\email{ted.sargent@northwestern.edu}), D. Sinton (\email{dave.sinton@utoronto.ca}), C.L. Zitnick (\email{zitnick@meta.com})}

\metadata[Code]{\url{https://github.com/FAIR-Chem/fairchem/tree/main/src/fairchem/applications/ocx}, MIT license}
\metadata[Dataset]{\url{https://fair-chem.github.io/core/datasets/ocx24.html}, Creative Commons 4.0 License}
\begin{document}

\maketitle

%%%MAIN TEXT%%%%
% \hline
%%%%% Introduction
\section{Introduction}

The climate perturbations and ecological instability resulting from fossil fuel combustion and industrial processes necessitates a shift towards renewable and carbon-neutral energy solutions. The electrochemical \gls{HER} holds the potential to generate green hydrogen from renewable energy for use in fuel cells, chemicals manufacturing, steel production and many other applications that could help abate \cd~emissions\cite{rasul2022future}. The electrochemical \gls{CO2RR} has emerged as an attractive method to transform our economy to one that is carbon-neutral through the recycling of \cd\cite{nitopi2019progress}. This includes the opportunity to offset the costs associated with carbon capture by creating sustainable value streams for \cd~utilization\cite{nitopi2019progress, smith2019pathways}. Currently, the deployment of \her~and \cdrr~are limited in part due to the need to find durable and low-cost catalysts that can drive chemical reactions efficiently and selectively toward valuable products\cite{smith2019pathways,verma2022cost}.

Catalyst discovery is a time consuming process of trial and error built on decades of expert domain knowledge\cite{tabor2018accelerating, FARRUSSENG2008487, senkan2001combinatorial, mccullough2020high}. Since the analysis of even a single material is time intensive, different areas of the material design space are typically evaluated in separate studies by independent institutions \cite{}. As is well documented \cite{scott2022err}, the lack of reproducibility of these studies leads to challenges in building upon prior results, and hinders our ability to draw conclusions from the studies in aggregation \cite{woldu2022electrochemical, yu2021recent}. 

Recently, the application of \gls{AI} to computational chemistry has dramatically increased the speed at which materials may be analyzed \cite{zitnick2020introduction,lan2023adsorbml,wander2024cattsunami,gnome,mattersim} as compared to traditional approaches using \gls{DFT}. Similar to other fields such as protein folding \cite{AlphaFold,esm} and large language models \cite{llama3,gpt4}, these ML approaches have improved significantly with large dataset initiatives including the Materials Project\cite{MP_main}, Open Catalyst Project\cite{chanussot2021open}, OQMD\cite{oqmd1, oqmd2}, and others\cite{omat24,spice, smith2020ani,trans1x}. However, computational approaches typically rely on idealized structures and properties, which may not accurately represent the real-world lab conditions. Furthermore, typical experimental material performance properties, such as product selectivity, cannot be directly estimated from commonly calculated computational properties such as adsorption or transition-state energies.

\begin{figure}
    \centering
    \includegraphics[width=0.95\linewidth]{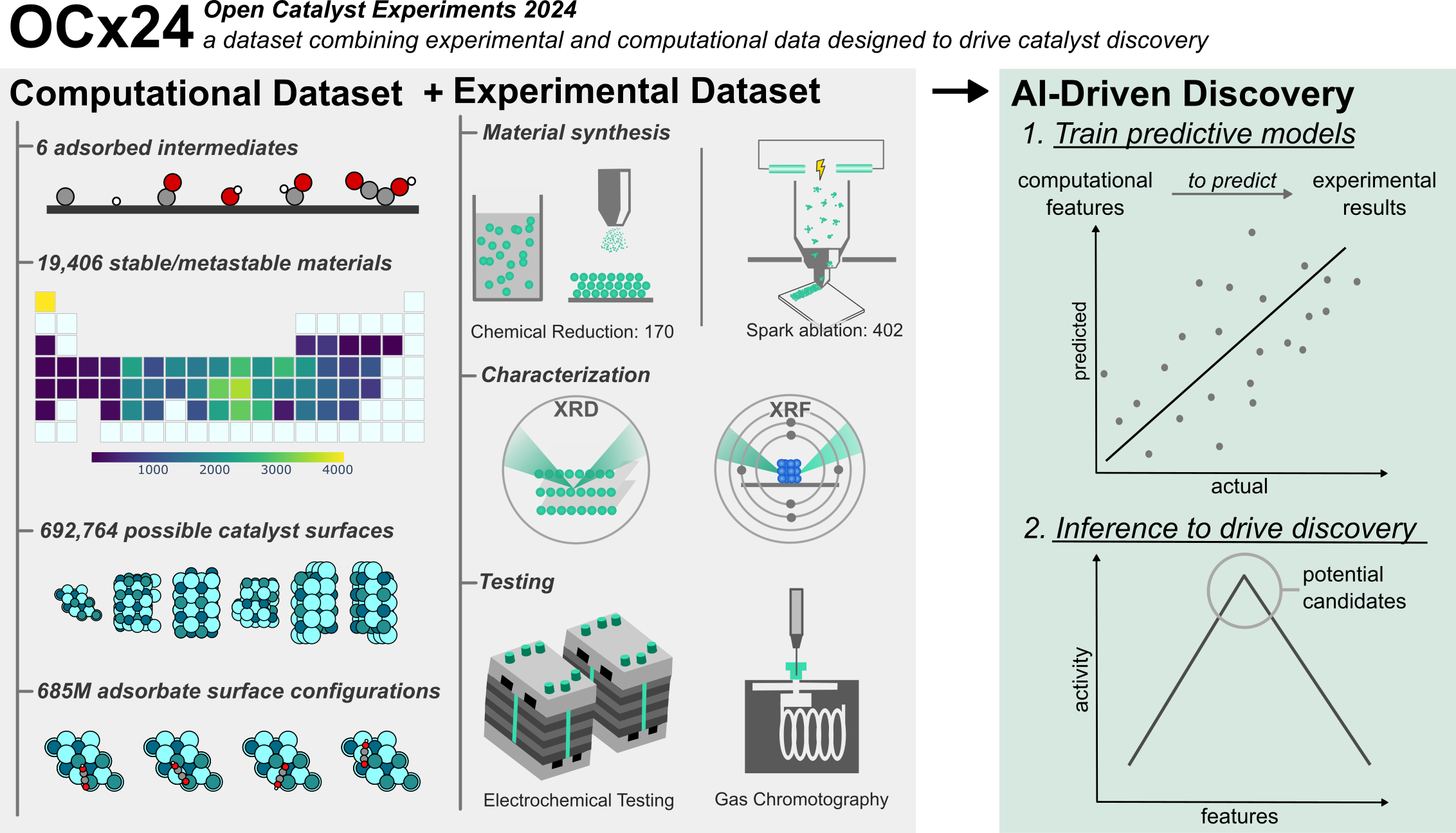}
    \caption[A summary of the work performed]{A summary of the computational and experimental screening efforts, and the resulting outcomes. Computationally, six adsorbates were used as descriptors and their adsorption energies were calculated across a wide swath of materials. Experimentally, two synthesis techniques were used to prepare 572 samples which were characterized and a subset tested. We used this data to build models capable of predicting experimental outcomes using computational features.}
    \label{fig:summary-fig}
\end{figure}

How can we bridge the gap between computational models and experimental results? In this paper, we introduce Open Catalyst Experiments 2024 (\ocx), in which we perform experimental studies specifically designed to aid AI models in mapping computational properties to experimental results, Figure \ref{fig:summary-fig}. To serve as effective training data, experimental studies must have several properties. The dataset should contain a diverse set of materials across the elemental space and cover both positive and negative results. Candidate samples should be evaluated in a reproducible manner and under industrially relevant conditions that may translate to large-scale reactors. Finally, the samples synthesized should be amenable to computational analysis. To accomplish this, we aim to study intermetallic nanoparticles\cite{interNPs}, which are structurally ordered metal alloys with precise atomic stoichiometry. They are ideal for modeling catalysts computationally as a result of their well-defined structure. However, the experimental synthesis of these intermetallic nanoparticles is exceptionally challenging\cite{montoya2024ai}. Synthesis conditions need to be carefully controlled to ensure the nanoparticles are of the appropriate size, single phase, and ordered\cite{cui2022multi}. Since high failure rates are common for synthesis, proper characterization is required for all samples to ensure they correspond to the target material. 

\begin{figure}
    \centering
    \includegraphics[width=0.82\linewidth]{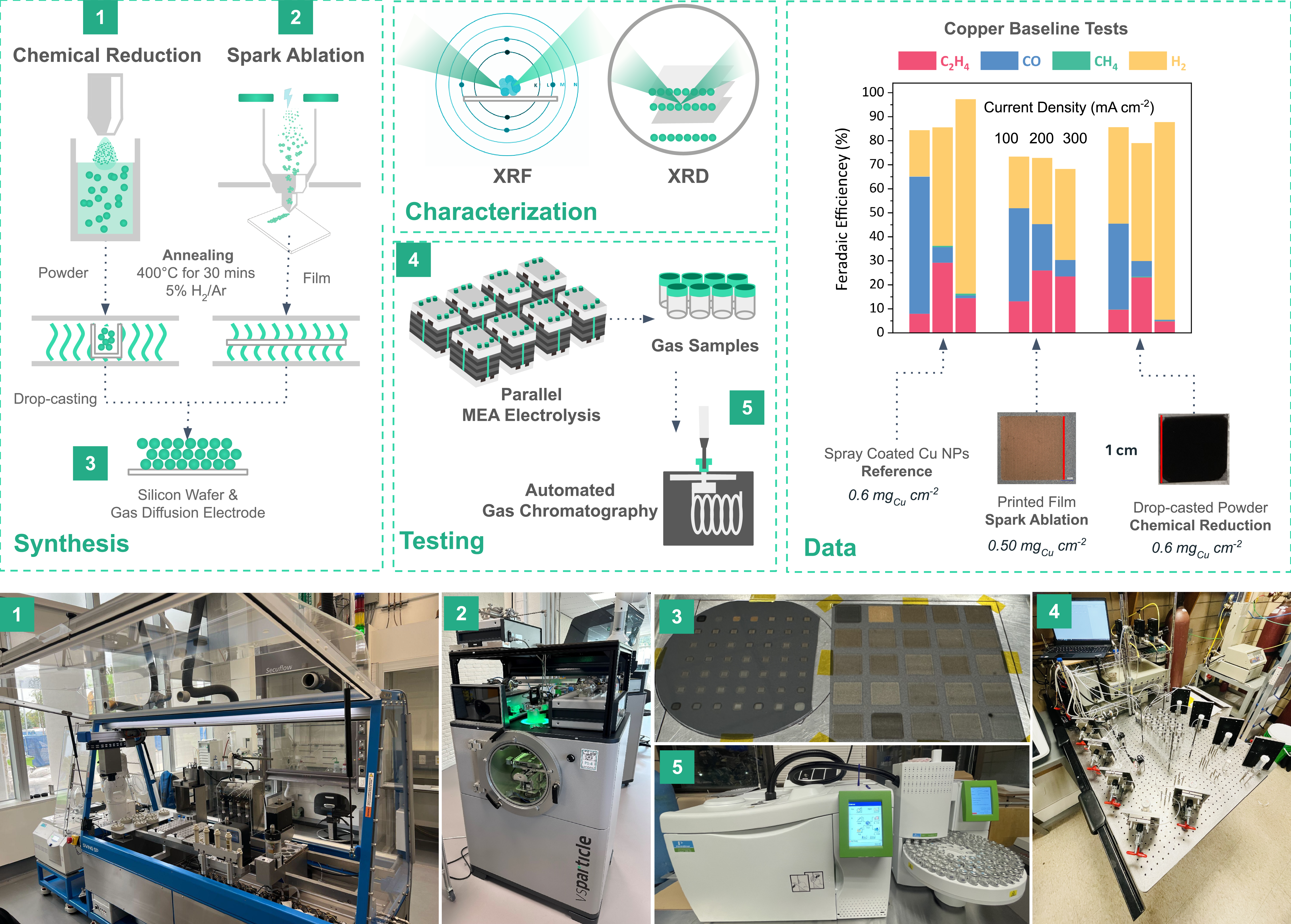}
    \caption{This figure illustrates the experimental pipeline, detailing the process from synthesis to characterization and testing. (right) Cu baselines for the two synthesis techniques and a reference technique using spray coated nanoparticles. Similar trends across current densities are seen for all techniques.}
    \label{fig:expt_pipeline}
\end{figure}

For experimental screening, \ocx~utilizes two synthesis techniques, chemical reduction and spark ablation, to synthesize 572  samples covering a diverse set of samples of 13 different elements, as shown in Figure \ref{fig:expt_pipeline}. To characterize each sample, \gls{XRF} and \gls{XRD} were performed to determine its composition and to elucidate the purity and structure of the sample synthesized, respectively. Using this information, we filtered the samples to those that were single phase and had good structural matches to the desired compositional structures as determined by an automated \gls{XRD} multiphase identification pipeline. Finally, electrochemical \cdrr~and \her~testing were performed at industrially relevant current densities with priority given to samples that passed the prior filtering step. Without replicates, there were 230 unique sample preparations, but there are samples that have very similar compositions. After aggregating similar compositions, 179 experimental targets remained for downstream modeling efforts. 

For computational screening, \ocx~calculates the adsorption energies of 6 adsorbate intermediates (OH, CO, CHO, C, COCOH, H) for 19,406 materials. We considered any material in the Materials Project (MP)\cite{MP_main}, Open Quantum Materials Database (OQMD) \cite{oqmd1, oqmd2}, and Alexandria\cite{dcgat, cgat} that could be thermodynamically stable under reaction conditions (Pourbaix decomposition energy less than 0.05 eV/atom). Adsorption energy calculations were performed on surface terminations up to Miller index two using the AdsorbML pipeline \cite{lan2023adsorbml} that combines AI and DFT calculations. This effort required 685 million structural relaxations and $\sim$20 million DFT single points and is the largest computational screening of catalysts for any application to date.

By combining the experimental and computational results, we built predictive models for \gls{HER}. For this reaction, platinum catalysts are known to perform well, but their high cost limits their commercial deployment~\cite{verma2022cost,hansen2021there}. Models were trained to predict the cell voltage at 50 mA/cm$^2$ production rate using the adsorption energies of H and OH as features. A linear model was used to perform inference on the full set of 19,406 stable/metastable materials. In this high-data regime, we recover a Sabatier volcano with Pt predicted to be near the apex, despite there being no Pt-containing alloys in the training dataset. Through this analysis, we identify hundreds of potential HER catalysts, many of which are importantly composed of low-cost elements. 

Finding a catalyst that outperforms copper for the production of multi-carbon products in \cdrr~remains an outstanding challenge\cite{chan2020few}. Despite copper's ability to produce multi-carbon products, improvements in activity, selectivity, and stability are desirable\cite{chan2020few, nitopi2019progress, zhou2020formation}. In our models, we predict \cdrr~production rates at a fixed applied potential to explore the selectivity of the catalyst. The correlation of the predictive models for H$_2$, CO, and liquid products is weaker than \her~for this more complex reaction. Building predictive models is especially challenging when the model must generalize to novel compositions outside the training dataset. Further advances in modeling and additional training data offer promising future directions. Although this paper focuses on \her~and \cdrr, we expect these findings to be useful for many applications and chemistries beyond those studied here.

%%%%% Methods
\section{Experimental Methods}

We determine the materials targeted for synthesis by sampling a diverse set of materials based on their elemental composition and a rough estimate of their expected products. For synthesis, we use two techniques to increase the diversity of the samples we use for testing, and to gain an understanding of the impact of different synthesis techniques. After characterization of the resulting samples using \gls{XRF} and \gls{XRD}, a subset of the samples is passed on for testing based in part on whether they match the target computational materials.

\subsection{Synthesis Techniques}

Diversity is essential when creating training data for ML models. When selecting materials to synthesize, this includes the diversity of their elemental composition and the diversity of the products produced when they are used as catalysts. To increase the likelihood of sampling materials that produce different products, we grouped the materials into three rough groups that are likely to produce hydrogen, C$_1$, and C$_{2+}$ products and attempted to sample from them equally for synthesis. See Section \ref{sec:SI_material-selection} for details on this process.

\begin{figure}
    \centering
    \includegraphics[width=1\linewidth]{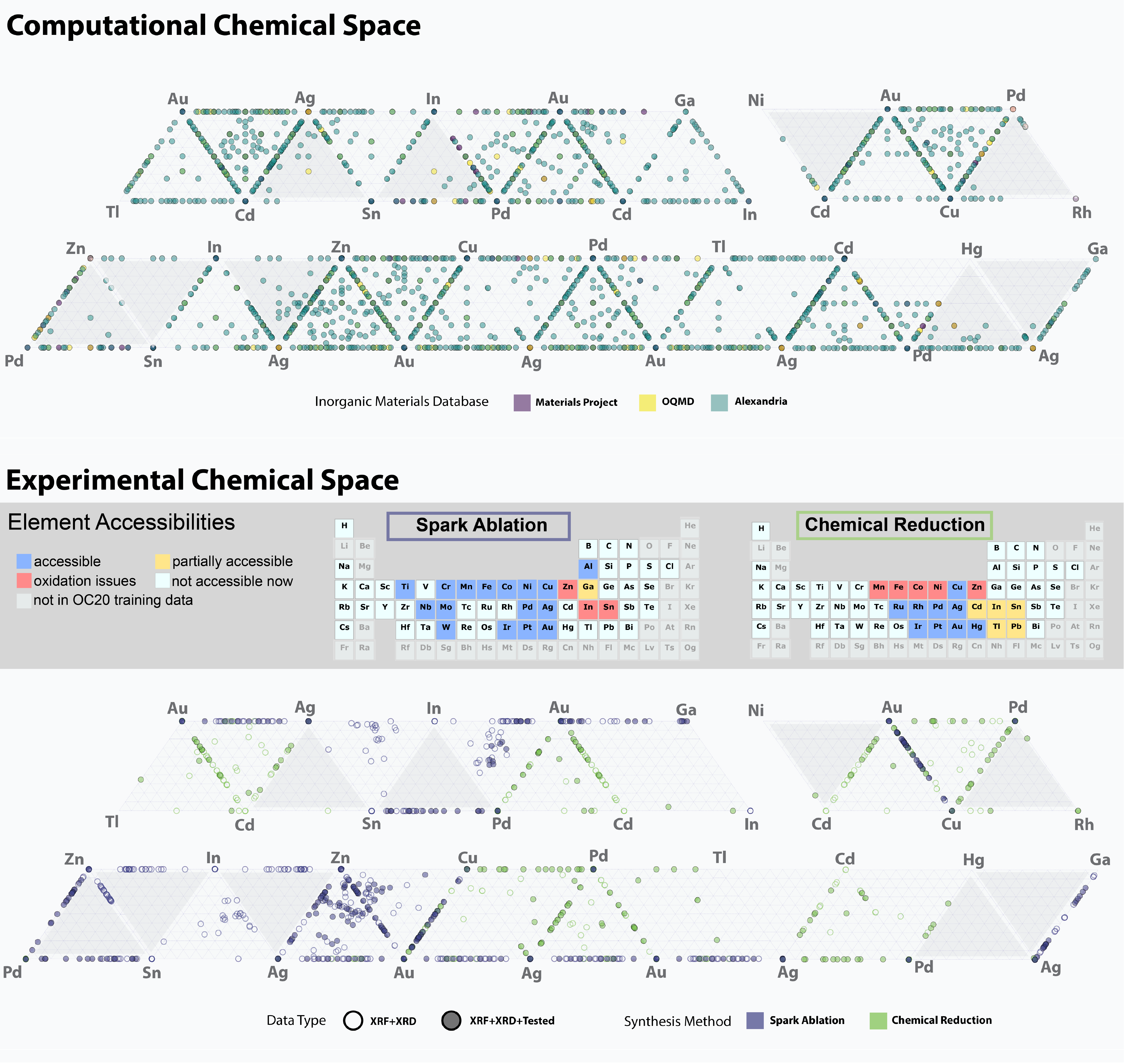}
    \caption{Chemical space anaylsis. (Top) Compositional diversity in the computational data considered for synthesis mapped and shown for each bulk database source: \gls{MP}, \gls{OQMD}, and Alexandria.  (Bottom) Depiction of the accessibility of elements via the two synthesis methods (chemical reduction and spark ablation), along with a chemical space representation for all experimentally synthesized, characterized, and tested samples. Darker shaded ternary phase diagrams result from connecting several diagrams and were intentionally not explored.}
    \label{fig:chemical_space}
\end{figure}

The synthesis of a diverse set of samples was the most significant challenge of this study. Given a desired composition, a synthesis technique should ideally be capable of mixing metals in nanoparticles with controlled size and mass loading. This helps to eliminate the influence of structural variance between different synthesized samples on the local reaction environment (i.e. thickness, porosity, inhomogeneity). Another challenge in synthesis is that the alloys need to be deposited on a carbon gas diffusion layer (GDL) to facilitate gas reactions for catalyst testing. Attempts to use high-temperature synthesis methods, essential for alloying more than one metal, such as carbothermal shock synthesis \cite{yao2018carbothermal} failed because the structure of the GDL was damaged during the process. Two synthesis techniques, chemical reduction and spark ablation, met our criteria and produced acceptable baseline results on Cu samples, see Figure \ref{fig:expt_pipeline} (right). We use both of these approaches to synthesize binary and ternary metal alloys in our study since they have complementary elemental accessibility, see Figure \ref{fig:chemical_space}. Duplicate samples were synthesized in a subset of the experiments to aid in verifying the reproducibility of the sample synthesis and testing procedures.

Spark ablation is a dry method, which uses additive printing with electrical energy to fragment metal rods into nanoclusters under an inert gas that can be directly deposited on a GDL. This was accomplished using a VSParticle© nanoparticle generator (VSP-P1). The generator relies on a physical phenomenon similar to vaporization and condensation, which uses high-voltage sparks between two closely spaced metal or alloy rods to generate a localized plasma. The intense heat from the plasma vaporizes material from these rods (ablation), which then cools and condenses into nanoparticles. Using an impaction technique and a 3D printer, nanoparticles can be synthesized and printed in a few hours. After printing a sample on the GDL, it is annealed using a tube furnace at 400\degree C temperatures with reducing gas to target intermetallic alloyed nanoparticles. See Section \ref{sec:SI_sparksynth} for details on this process.

Chemical reduction was performed using an automated robotic system (Chemspeed) that performs wet chemistry operations. This approach uses metal salt precursors that are mixed and reduced using a reducing agent that donates electrons to the mixed metals. The resulting sample is then post-annealed at elevated temperatures to create alloyed nanoparticle powders. These powders are combined with a polymer binder and applied to a GDL via drop casting for testing. See Section \ref{sec:SI_chemreduction} for more details.

\subsection{Experimental Testing}

Parallelized screening for \cdrr~and \her~was performed using a modified \gls{MEA} setup from a previous study \cite{kose2022high}. The testing process began with the electrolyzer operating at 50 mA/cm$^2$ for 9 minutes to evaluate \her~activity. Afterward, CO$_2$ gas was introduced (flow rate: 30 s.c.c.m.) into the electrolyzers, and the \cdrr~was carried out at current densities of 50, 100, 150, 200, and 300 mA/cm$^2$ for 9 minutes each. The gas products were collected into headspace vials during each step of the \cdrr~process and subsequently analyzed using gas chromatography (PerkinElmer Clarus 590) with an autosampler (PerkinElmer TurboMatrix HS110). The gas products were detected by a thermal conductivity detector (TCD) and a flame ionization detector (FID) using high-purity Ar ($99.999\%$) as the carrier gas. For further details, see Section \ref{sec:SI_testing}.

\begin{figure}[h!]
    \centering
    \includegraphics[width=0.8\linewidth]{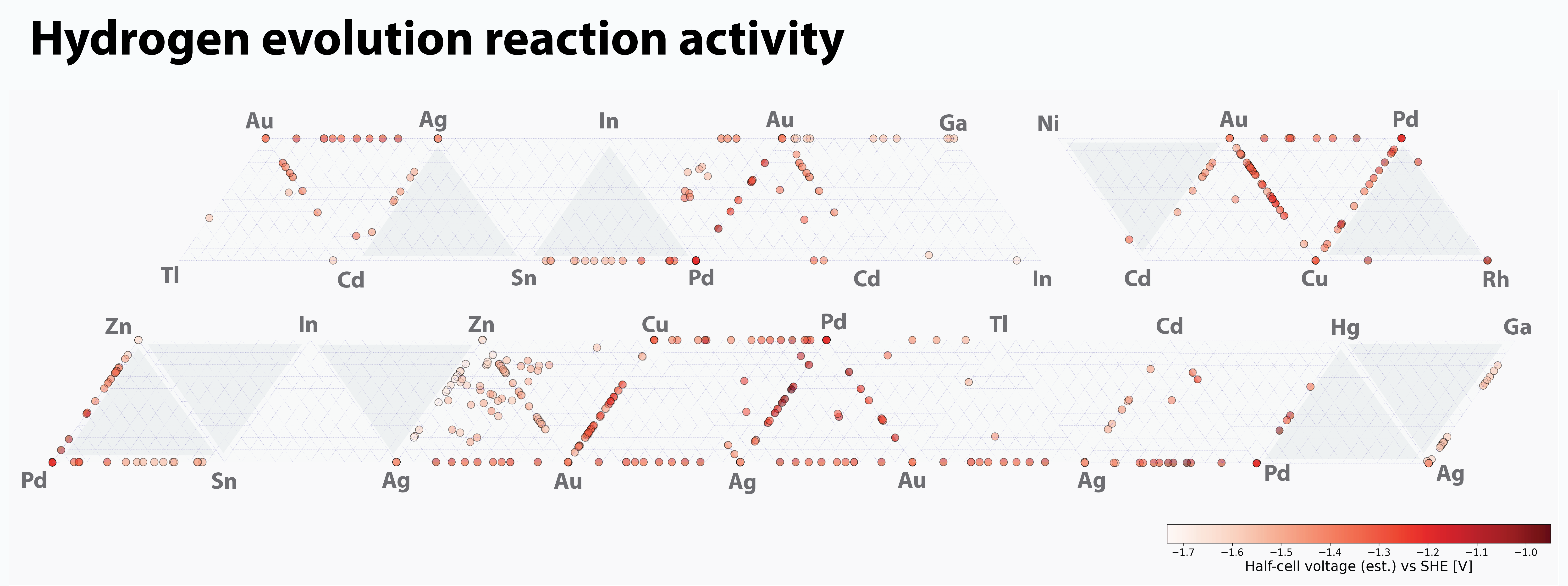}
    \caption{Mapping of half-cell cathodic potential vs SHE for hydrogen evolution reaction Darker
shaded ternary phase diagrams result from connecting several diagrams and were not intentionally explored}
    \label{fig:h2voltage_space}
\end{figure}

For the experimental results of \her, we recorded the full cell voltage of the \gls{MEA}. Using a three-electrode \gls{MEA} cell setup, we measured voltage losses across each component—membrane, anode, and internal resistance (IR) losses to identify voltage losses. This method, validated in a previous study \cite{fatemah_unpublished_work} with the same type of \gls{MEA} cells used in this work, allowed us to estimate the cathodic half-cell potential relative to the \gls{SHE} (see Figure \ref{fig:h2voltage_space}. Ideally for \cdrr~we would record production rates for different products at a fixed voltage. However, our studies used five fixed current densities. To account for this, we use logarithmic linear interpolation to estimate results for a fixed voltage, see Section \ref{sec:SI_fixedpotential}. The liquid product Faraday efficiency was estimated by subtracting the total percentage of the gaseous product from 100\%.

\section{Characterization Methods}

\subsection{Composition and Structure Validation}

After synthesis, characterization is needed to verify whether the sample produced matches the target material. Precise verification of the target material is challenging and difficult to perform at high throughput \cite{hattrick2016perspective}. In our study, we use two techniques: \acrfull{XRF} and \acrfull{XRD}. \gls{XRF} measures the composition of the sample, that is, the percentage of elements in the sample. \gls{XRD} provides insights into the fraction of material phases existing in the sample and the crystal structure of each phase. It is important to note that a sample that matches both \gls{XRF} and \gls{XRD} measurements does not guarantee that it is the same as the target material, since both measurements possess inherent limitations. \gls{XRF} provides elemental analysis without specifying chemical forms, while \gls{XRD} may encounter challenges with similar crystal structures, complex phases, and low crystallinity. Although integrating \gls{XRF} and \gls{XRD} enhances material identification, ambiguities and detection limits persist, which necessitates careful interpretation and potentially supplementary methods. However, they are effective in finding true negatives and samples that contain multiple phases. All characterization measurements were performed on samples deposited on silicon wafers to avoid an interference \gls{XRD} signal from the underlying GDL substrate. For the spark ablation synthesis, the GDL and silicon wafer samples were printed at the same time, one layer at a time, to ensure the samples closely matched. See Section \ref{sec:SI_characterization} for \gls{XRF} and \gls{XRD} details.

\subsection{XRD Analysis}
To conduct \gls{XRD} analysis and identify phase combinations in samples, we developed an automated tool that streamlines the matching and refinement process. This tool compares experimental \gls{XRD} pattern, i.e., peak position and intensity array, with computational \gls{XRD} patterns that are generated from structures in the Materials Project, OQMD, and Alexandria, as well as experimental references from the Crystallography Open Database (COD) \cite{Grazulis2012}. The process is conducted in four steps:
\begin{itemize}
    \item Similarity Analysis (Figure \ref{fig:xrd_step1}): We start by matching the position and intensity of experimental \gls{XRD} pattern with simulated patterns from computational structures.
    \item Multiphase matching (Figure \ref{fig:xrd_step2}): Once a phase is matched, we remove it from the experimental \gls{XRD} pattern. We then repeat the first step to find additional phases. This process continues until all phases are identified or a user-defined limit is reached.
    \item Matching goodness evaluation (Figure \ref{fig:xrd_step3}): Finally, we refine the XRD fit using Rietveld refinement\cite{mccusker1999rietveld, baptista2022xerus}. This step checks how well the identified phases match the experimental data. A quick and rough Rietveld fitting helps us assess the quality of the match by calculating the weighted profile R score (R$_{wp}$), which evaluates the fit’s quality (the lower, the better). We report fits that are below an $R_{wp}$ threshold of 50, as larger values typically indicate poor-quality fits that do not provide additional value to our analysis.
\end{itemize}

To find the best crystal structure match to the synthesized sample, we look for XRD matching solutions (phase mixtures) containing one phase that meet the following two conditions:
\begin{itemize}
    \item The major phase has a weight fraction more than 70 wt.\%  of the phase mixture.
    \item The composition (atomic fraction) in the matched phase is close to what we measured in the sample using \gls{XRF} (within $\pm$ two standard deviations). We determine the standard deviation by measuring the \gls{XRF} composition at three different spots on the sample. 
\end{itemize}
We then rank the phases that meet the above criteria based on the refinement $R_{wp}$ score. We consider the phase to be a potential match using a score we call $q$, which is calculated as follows:
\begin{equation}
q_i = 100 \cdot \exp(-\alpha \cdot (R_{wp,i} - R_{wp,min}))
\end{equation}
In this equation, $\alpha$ is a constant scaling factor set to 0.05. $R_{wp,i}$ is the $R_{wp}$ value of the phase being considered and $R_{wp,min}$ is the lowest $R_{wp}$ value (best fit) across all of the phases considered as potential possibilities to match the measured \gls{XRD} spectrum. The $q$-score is used to assess the goodness of fit relative to other possible solutions. In this work, we designate samples as matched if they have a phase with weight fraction $\geq$70 wt.\%, $q_{i} \geq 70$ and an R$_{wp} \leq 40$. See Section \ref{sec:SI_characterization} for more details.
\section{Computational Methods}

Our goal is to develop computational models capable of mapping a set of descriptors for a material to their observed experimental performance over a broad range of materials. Other attempts at computational modeling use experimental data mined from literature \cite{suzuki2019statistical, mine2021analysis, mine2022machine}, which is significantly more challenging since the materials may be synthesized and tested under varying conditions, unlike this study. Using results from single lab studies simplifies the problem. Without consistent testing conditions, experimental conditions along with elemental features have been used as inputs to ML models for guiding the selection of performant materials \cite{williams2019enabling, wang2021high, suvarna2024active}. 

In our study, we explore the use of \textit{ab initio} features in predicting experimental outcomes. \textit{Ab initio} features have been shown to be predictive of catalyst trends\cite{kulkarni2018understanding, norskov2005trends, latimer2017understanding} and are used to computationally screen material spaces\cite{wander2022catlas, tran2022screening,2d_mat_HER, SA_W_ML, SA-MoS2, DA-NF}. There are only a few examples where screening is successfully used to identify catalyst candidates that are experimentally performant\cite{greeley2006computational, zhong2020accelerated}. \ocx~goes beyond what has previously been performed by focusing on material diversity and their performance in complex reactions like \cdrr.

\subsection{Computational Material Selection}

As inputs to our computational models, we consider all materials available in three permissively licensed computational materials databases: \gls{MP}~\cite{MP_main}, \gls{OQMD}~\cite{oqmd1, oqmd2}, and Alexandria~\cite{dcgat, cgat}. These datasets cover a much wider range of materials than our experimental dataset and may be useful in guiding future material exploration. In an effort to improve the likelihood of discovering an industrially viable catalyst, we filtered materials by assessing their thermodynamic stability under reaction conditions. To do this, we evaluate the decomposition energy using the Pourbaix framework implemented in pymatgen\cite{pbx_efficient, pymatgen, pbx_main}, see Section \ref{sec:compmatselection} for details. This results in 19,406 total materials after deduplication (2,458 materials from \gls{MP}, 3,173 materials from \gls{OQMD}, and 13,775 materials from Alexandria) for use in our computational pipeline. 

\subsection{Descriptor Selection}

\textit{Ab initio} features based on adsorption energies, which estimate the attraction of different intermediate molecules to the surface of a catalyst, are commonly used as descriptors. Bagger et al.\cite{bagger2017electrochemical} calculated the adsorption energies of five intermediates (H, CO, COOH, CH$_3$O and HCOO) for 16 unary materials with known product preferences to determine which descriptors provide separability of products in the descriptor space. They recommend using H, COOH, CO and an adsorbate with oxygen binding character to gain full separability. Peng et al.\cite{peng2021role} show that monoatomic carbon is an important intermediate for the production of multi-carbon products and use C and CO to make a selectivity map to search for alloy candidates. Liu et al.\cite{liu2017understanding} make a selectivity heatmap using CO and the H-CO transition state as descriptor energies. Later, Zhong et al.\cite{zhong2020accelerated} demonstrated that the same character could be retained using H and CO adsorbates.

We selected our list of adsorbates (C, H, CO, OH, CHO, and COCOH) to include the characteristics and descriptors identified in these prior works with the objective of designing universal predictive models. In addition to the simple adsorbates, CHO and COCOH mark key points in the reaction network to produce non-formate C$_1$ and C$_{2+}$ products \cite{nitopi2019progress}, while OH offers orthogonal information since oxygen species generally do not scale linearly with hydrogen and carbon. All adsorbates are shown in Figure \ref{fig:summary-fig}. For each adsorbate, the adsorption energy is calculated using the hybrid ML and DFT AdsorbML \cite{lan2023adsorbml} pipeline for all Miller indices up to two for the entire dataset of 19,406 materials. These calculations required more than 685 million structural relaxations and $\sim$20 million DFT single points and is the largest computational screening of catalysts for any application to date. See Section \ref{sec:adsorbcalc} for additional details.   

In addition to \textit{ab initio} features, we also consider descriptors for bulk material properties using Matminer\cite{ward2018matminer}. In particular, we included the material Mendeleev number, row number, group number, atomic mass, atomic radius, and Pauling electronegativity. This includes characteristics that do not have an obvious relationship to catalytic properties.

\subsection{Predictive Models}

Catalyst performance for nanoparticles is an aggregate property over many surfaces, which may or may not be catalytically active. A predictive model may combine surface-level descriptors (e.g., adsorption energies) with bulk material properties (e.g., Matminer features) to estimate the experimental performance of the aggregated material. It is clear how material-level properties can be leveraged to predict other material-level properties, but aggregating surface-level properties to predict material-level properties is more challenging. This is in part due to the fact that the estimation of which surfaces will be realized experimentally is a difficult \cite{tao2011situ, zafeiratos2012alloys} and open problem. We explore three approaches to combining surface-level descriptors. First, we consider the simplest case - using the mean adsorption energy across all surfaces:
\begin{equation}
    E_{ads,mean} = \frac{1}{N}\sum_i^N E_{ads,i}
\end{equation}
where $i$ iterates over the $N$ surfaces and $E_{ads,i}$ is the surface's adsorption energy. The second uses Boltzmann weighting for both the adsorption energies and the surface energies given the set of surfaces using:
\begin{equation}
    E_{ads,Boltz.} = \frac{\sum_i^N E_{ads,i} e^{-(E_{ads,i} - E_{ads,min})/(k_b T)} e^{-(E_{cl,i} - E_{cl,min})/(k_b T)}}{\sum_i^N   e^{-(E_{ads,i} - E_{ads,min})/(k_b T)} e^{-(E_{cl,i} - E_{cl,min})/(k_b T)}}
\end{equation}
where $E_{cl,i}$ is the cleavage energy on that surface and $E_{ads,min}$ is the material's minimum adsorption energy and similarly for $E_{cl,min}$. This results in surfaces with lower energies, which are more likely to be realized in practice, having higher weight. It also results in lower, more favorable adsorption energies having higher weight.  The third approach uses the Wulff construction\cite{ringe2011wulff}, which is a minimization of the total energy over a nanoparticle as a function of the surface energies of the various possible terminations with different geometric character\cite{dobrushin1992wulff}. For non-unary materials, it can be difficult to calculate surface energies, but the cleavage energy may be used as a proxy. Wulff constructions were performed using pymatgen\cite{pymatgen, tran2016surface}. Given a subset of $M\subset N$ surfaces appearing on the Wulff shape for a material with facet fractions $w_i, i\in M$ ($\sum_i w_i = 1$), we calculate the Wulff weighted energies using:
\begin{equation}
    E_{ads,Wulff} = \sum_i^M w_iE_{ads,i}
\end{equation}
Finally, the experimental values $x$ are predicted using:
\begin{equation}
    x =  f(E_{ads,X}, b)
\end{equation}
where $f$ is a function learned from the experimental training data, and $E_{ads,X}$ is one of the aggregated adsorption energies (mean, Boltz., Wulff), and $b$ are the bulk descriptors (e.g., Matminer features). When cleavage energies were not available (convergence errors, etc.) to compute Wulff and Boltzmann weighted energies, mean energies were used. All features were standardized to unit variance before training.

We employ two regression models that will be presented here: a simple linear regression model and a random forest regression model. Besides these models, we also explored Gaussian process, kernel ridge, support vector regression, and gradient boosting regression models; all use the implementation in scikit-learn\cite{scikit-learn, sklearn_api} and xgboost\cite{chen2016xgboost} Python packages. 

%%%%% Results 
\section{Results}
For our results, we explore whether models trained on experimental data using computational descriptors as input can produce results that are predictive of held out experimental data. Given the chemical diversity of the data, and the complexity of reactions like \cdrr, we are mainly looking at whether the models can predict trends and we are not focused on the absolute accuracy of the results. In many applications, the ability to rank candidates for a particular reaction is itself an impactful contribution. It is our hope that this study will provide the basis for future research in further understanding the connections between computational models and experimental studies. 

For each model, we evaluate its performance under two different \gls{CV} strategies:
\begin{itemize}
    \item[] \textbf{\gls{LOO}}: Hold out a given sample, train on the remaining data, and evaluate on the held out sample.
    \item[] \textbf{\gls{LOCO}}: Hold out all samples with a given composition (i.e. all Au$_x$Cu$_y$ samples, where x,y $>$ 0), train on the remaining data, and evaluate on all held out samples.
\end{itemize}
A high correlation for \gls{LOO} indicates the model can predict outcomes for samples that share compositions with those in the training dataset, while a high \gls{LOCO} correlation indicates the model can generalize to novel compositions. \gls{LOCO} presents a more difficult task for the model, removing the ability to interpolate between samples of similar composition. 

\begin{figure}[h!]
    \centering
    \includegraphics[width=0.8\linewidth]{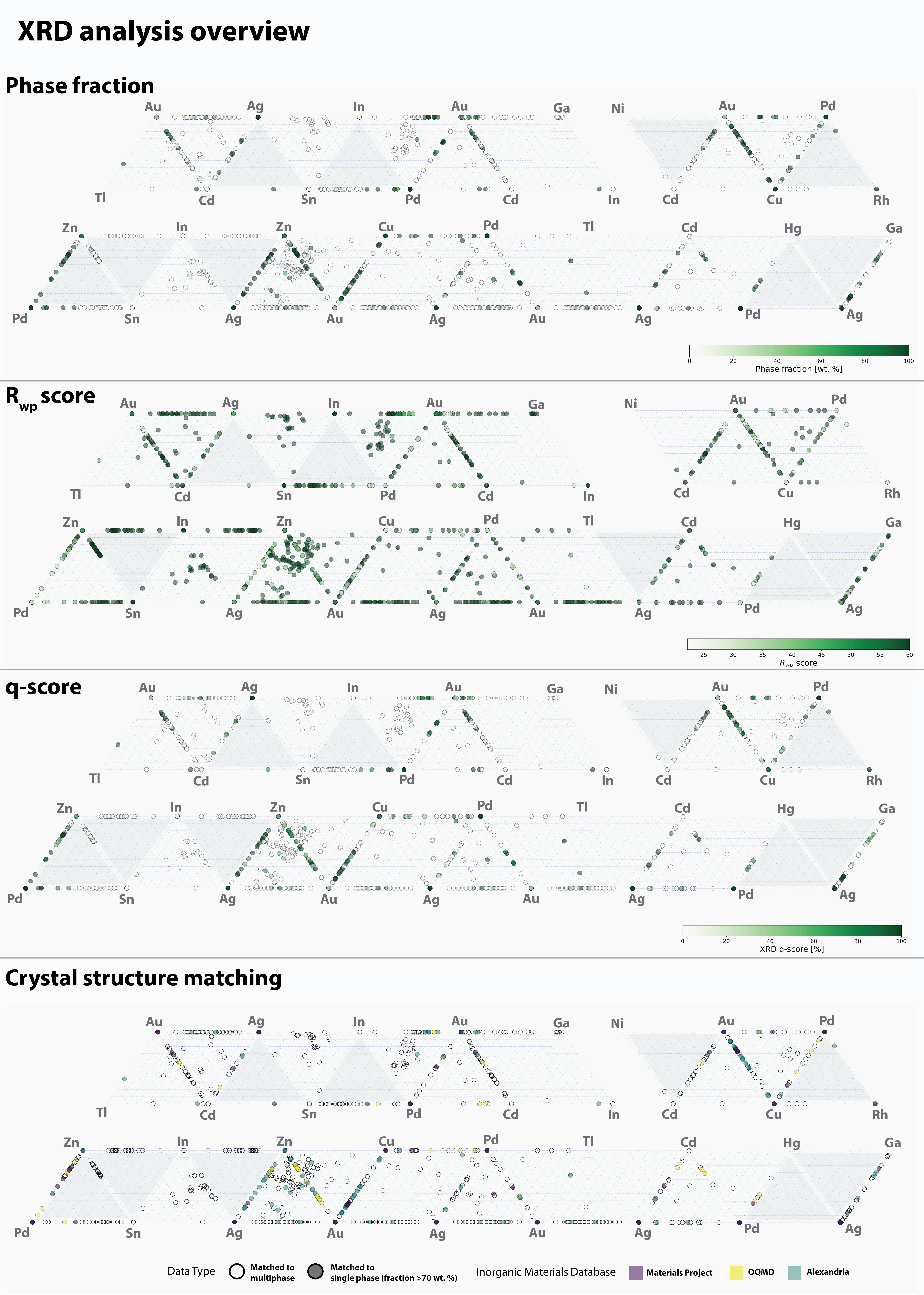}
    \caption{Mapping of XRD analysis results including phase fractions, $R_{wp} score$, $q-score$, and crystal matching indicating cif file source. Darker
shaded ternary phase diagrams result from connecting several diagrams and were not intentionally explored}
    \label{fig:xrd_space}
\end{figure}

The dataset contains samples that are matched and unmatched to an exact crystallographic structure (see Figure \ref{fig:xrd_space} for XRD analysis results). Matching was performed using the characterization data from \gls{XRD} and \gls{XRF}. When a match was available, we aggregated adsorption energies from only the corresponding matched bulk. Otherwise, we aggregated across the bulk material or materials with a composition closest to that of the \gls{XRF} composition. Of the 441 electrochemical tests performed, 230 were intended to be distinct sample preparations. Among the 230, there were samples with compositions that were very close to each other. We averaged over these samples to obtain 179 experimental targets for modeling efforts, 43 of which were matched (Phases with a fraction $\geq$70 wt.\%, $R_{wp}$ score $\leq$40 and q-score $\geq$70\%). We found that training jointly on matched and unmatched samples, outperforms just training on matched alone (Figure \ref{fig:match}). The complete set of tabulated results for our best performing \her~and \cdrr~models are provided in Table \ref{tab:full-results}. 

There are several nuances to handling the data that are worth noting. In some cases, samples labeled as matched have similar stoichiometries and match to the same bulk material. In this case, the sample with the composition closest to its matched bulk composition was retained and others discarded. For unmatched samples, if multiple samples correspond to the same bulk material based on XRF, the experimental results were averaged across all samples with the same synthesis technique. Lastly, if any sample in a group of replicates contained an \gls{XRD} match, the result is also considered a match. Across replicates, we computed the mean experimental measured quantities. Alternative methods to aggregating experimental samples may be explored for downstream predictive tasks.

\subsection{Hydrogen Evolution Reaction (\her)}
\begin{figure}[ht]
    \centering
    \includegraphics[width=1.0\linewidth]{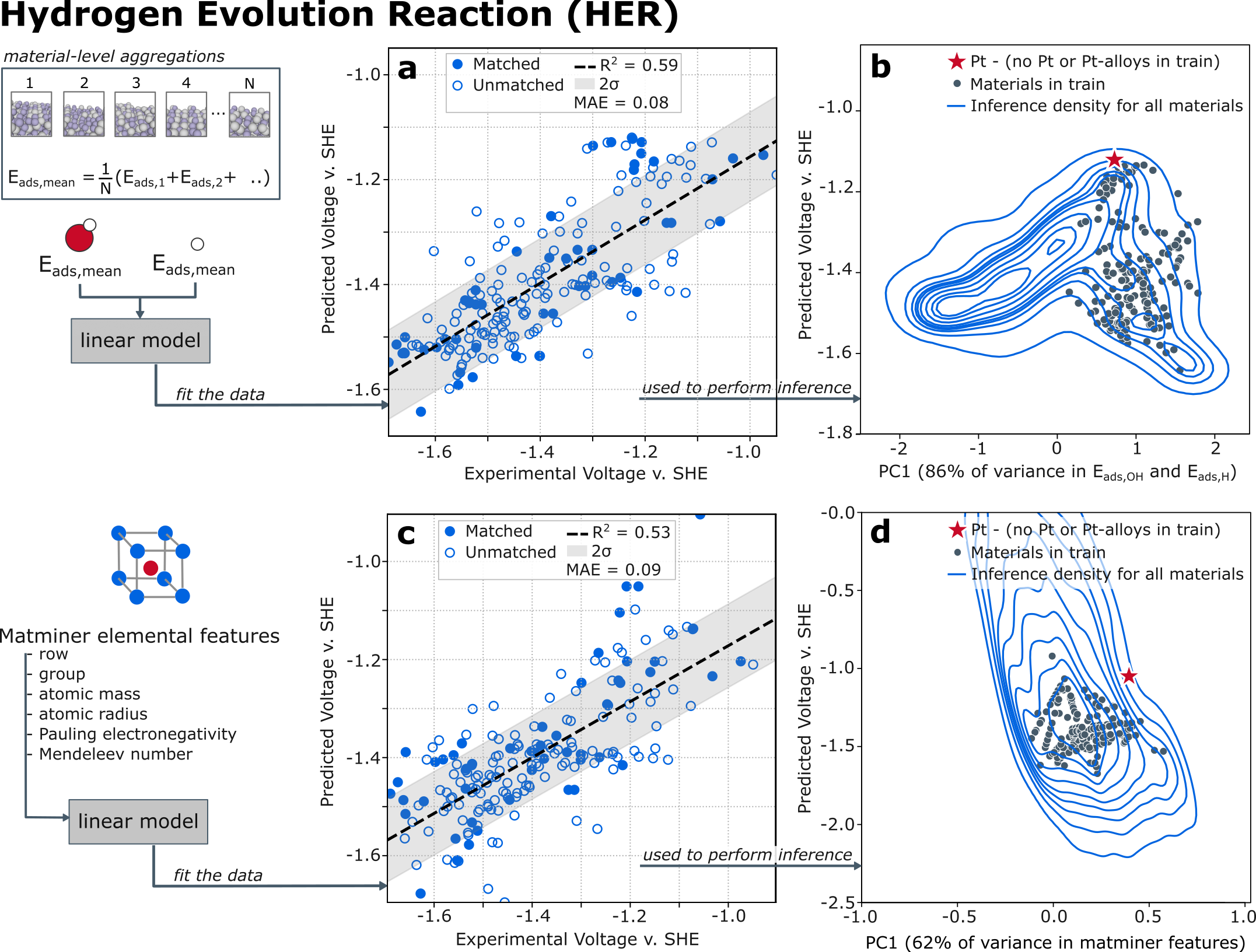}
    \caption{A summary of the inference campaign using models fit on hybrid data with (top) mean OH and H energies as features and (bottom) using Matminer features. A simple, linear model was trained with \gls{LOCO} \gls{CV} to obtain the predictive models (a and c). Using this fit and the computational descriptors for all stable/metastable materials assessed, we performed inference (b and d). The model trained on Matminer features does not generalize well (d), but the model trained on adsorption energies reveals a data driven Sabatier plot (b). Pt, which is annotated at the top of the volcano, was completely absent from the training data - even as an alloy.}
    \label{fig:her-inference}
\end{figure}

The results of training models to predict the experimental voltage for \her~is shown in Figure \ref{fig:her-inference}. Results are shown for a linear model trained on (top) H and OH adsorption energies as features and (bottom) Matminer features. Additional results can be found in Section \ref{sec:SI_herresults}. A parity plot between the predicted voltage and experimental voltage using \gls{LOCO} cross validation is shown in Figure \ref{fig:her-inference}a and \ref{fig:her-inference}c. The predictions and actual values are fairly well-correlated for both feature sets with R$^2$ = 0.59 for adsorption energies and R$^2$ = 0.53 for Matminer features. The trendline for the correlation is shown in Fig. \ref{fig:her-inference}a and \ref{fig:her-inference}c as a black dashed line. There was a modest performance improvement (R$^2$ = 0.61) by including all adsorption energies rather than just OH and H, but this result is not shown here for simplicity. On average, we observe an experimental standard deviation of $\sigma=0.043$V vs SHE for 297 sample replicates within $\pm$5\% in \gls{XRF} composition, an estimate of the noise in our targets (shaded gray in Figure \ref{fig:her-inference}).

Using these linear models, we performed inference across all 19,406 stable and metastable materials included in the computational workflow, which is shown in Fig. \ref{fig:her-inference}b and \ref{fig:her-inference}d. A simple \gls{PCA} was performed to collapse the features into a single dimension (x-axis) for the purposes of visualization. The principal component found explains 86\% of the variance for the adsorption energy features and 62\% for the Matminer features. The y-axis (predicted voltage v. SHE) should be taken to be a reflection of the material activity, with more positive values being more desirable. The samples appearing in train are annotated individually as points on the plot. Platinum, which does not appear in the training data and is well known as an effective catalyst for \her, is annotated by a red star. The inference data is represented as a density contour plot using kernel density information since the number of materials considered is so large. In this high-density data regime, we recover a Sabatier volcano for the model trained on mean OH and H adsorption energies, with platinum sitting at its peak despite platinum not appearing in the training data even as an alloy. Platinum is accessible by both of our synthesis methods, but platinum-containing materials were not selected as a part of our material selection process in part because of random chance.

Despite the models trained on Matminer\cite{ward2018matminer} features achieving similar correlations to those trained on adsorption energies, when using Matminer features for inference we do not predict platinum to be an apex catalyst as shown in Figure \ref{fig:her-inference}d. In fact, roughly half of all materials assessed are predicted to be more active than platinum (less negative voltage v. SHE). This is likely because the adsorption energies are more generalizable, underscoring their usefulness.

There are 3,869 materials that are predicted to have a voltage within twice the \gls{MAE} of Pt or better. A complete list of the 3,869 materials has been included with the datasets. Among the materials predicted to be good \her~catalysts, there are 19 Mo-S, Mo-Se, and Mo-Se-S alloys. This aligns well with experimental precedent that these materials are good \her~catalysts\cite{morales2014amorphous, yan2014recent, benck2014catalyzing, kwon2020se}. There are also 2,447 materials that contain Pd and / or Pt. This is not surprising since unary Pt and Pd are high-performing elemental catalysts for \her\cite{hori1994electrocatalytic, kuhl2014electrocatalytic}. Still, there are 436 materials that do not contain Pd, Pt, Rh, Ir, Ru, Au, Os, or Ag which are potential low-cost \her~catalysts. Among these, 282 are Se containing. The 200 best potential low-cost catalysts have been included in a table in the supplementary information, Section \ref{sec:HER_candidates}.

\begin{figure}
  \begin{minipage}[b]{0.55\linewidth}
    \begin{tabular}{l|cc|cc}
\multicolumn{1}{c}{} &  \multicolumn{4}{c}{$\mathrm{R^2}$} \\ %\midrule 
\multicolumn{1}{c}{} & \multicolumn{2}{c}{\her} & \multicolumn{2}{c}{\cdrr} \\ \midrule
Features & LOO & LOCO & LOO & LOCO   \\ \midrule
$\mathrm{E_{ads,mean}}$              & 0.61          & \textbf{0.59} & 0.29          & 0.05 \\
$\mathrm{E_{ads,Wulff}}$             & 0.44          & 0.41          & 0.03          & 0.00 \\
$\mathrm{E_{ads,Boltz.}}$            & 0.26          & 0.21          & 0.10          & 0.00 \\
Matminer-only                        & 0.58          & 0.52          & 0.39          & 0.10 \\
$\mathrm{E_{ads,mean}}$ + Matminer   & \textbf{0.63} & \textbf{0.59}          & \textbf{0.46} & 0.21 \\
$\mathrm{E_{ads,Wulff}}$ + Matminer  & 0.59          & 0.54          & 0.34          & 0.16 \\
$\mathrm{E_{ads,Boltz.}}$ + Matminer & 0.59          & 0.54          & \textbf{0.46}          & \textbf{0.22} \\
\toprule
\end{tabular}
\captionof{table}{Feature analysis on \her~and \cdrr~for their predictive performance. Results evaluated under both cross-validation strategies - \gls{LOO} and \gls{LOCO}. $\mathrm{H_2}$ was used for \cdrr~correlations. Adsorption energy features have a strong correlation for \her~but little to no correlation for \cdrr.}
\label{tab:features}
  \end{minipage}\hfill
  \begin{minipage}[b]{.38\linewidth}
    \centering
    \includegraphics[width=\linewidth]{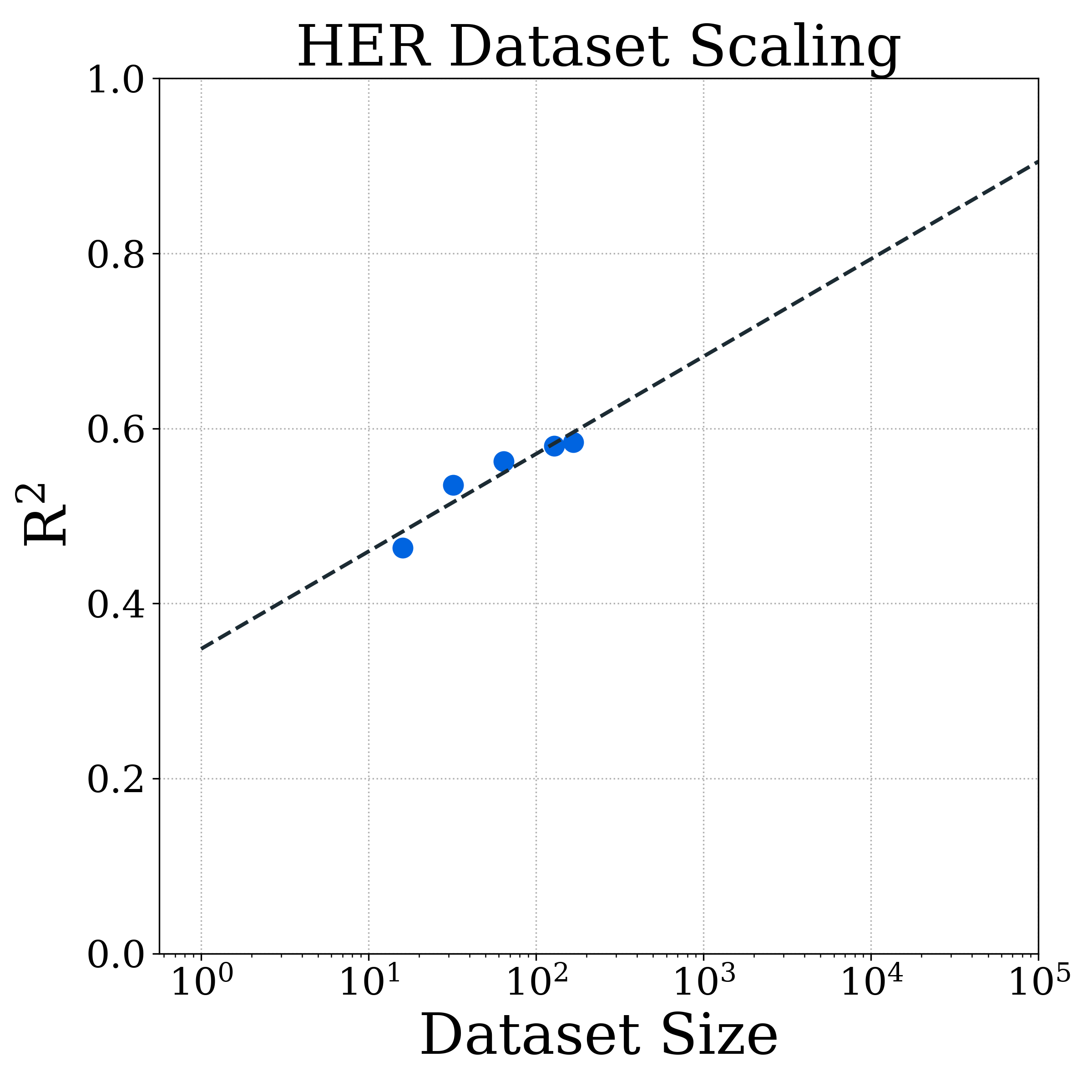}
    \captionof{figure}{\her~correlation results as a function of dataset size evaluated using \gls{LOCO} cross-validation.}
    \label{fig:scaling}
  \end{minipage}
\end{figure}

The correlation between experimental results and computation descriptors improves with training dataset size, as explored in Figure \ref{fig:scaling}. From this analysis we project that increasing the dataset size to 10$^4$ or 10$^5$ will allow for significantly more predictive models to be built.

\begin{figure}[h]
    \centering
    \includegraphics[width=0.95\linewidth]{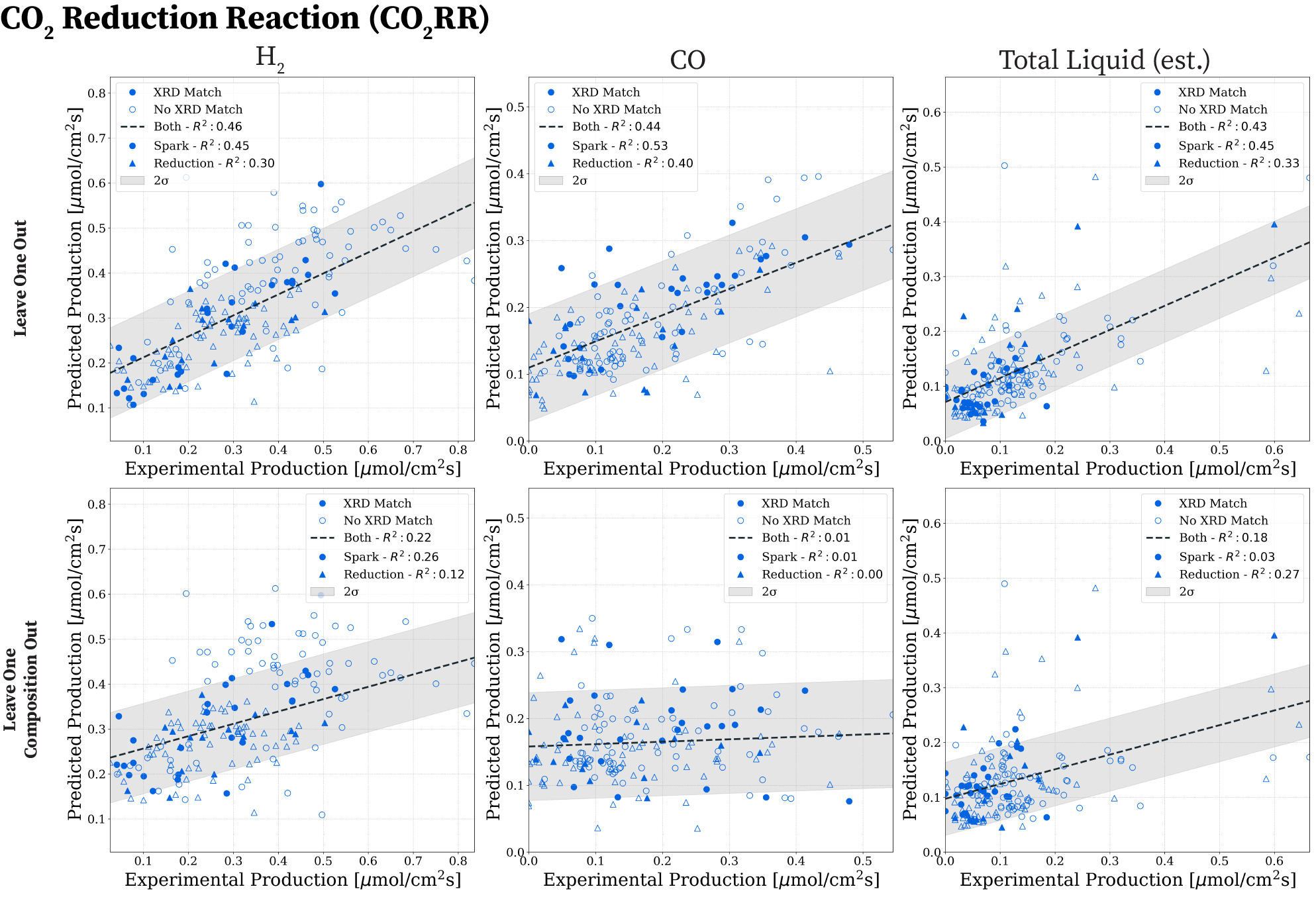}
    \caption{A summary of the \gls{CO2RR} predicted production rates across several products - $\mathrm{H_2}$, CO, and Total Liquid (estimated) for the two different cross-validation strategies. Results showcase the complexity when trying to predict on unseen compositions (\gls{LOCO} - bottom).  A random forest regression model was used with features coming from Boltzmann weighted adsorption energies and elemental Matminer features. The shaded region corresponds to twice the average standard deviation in the experimental targets for samples within $\pm$5\% \gls{XRF} composition.}
    \label{fig:co2rr-main}
\end{figure}

\subsection{\cd~Reduction Reaction (\cdrr)}

\begin{figure}[h!]
    \centering
    \includegraphics[width=0.8\linewidth]{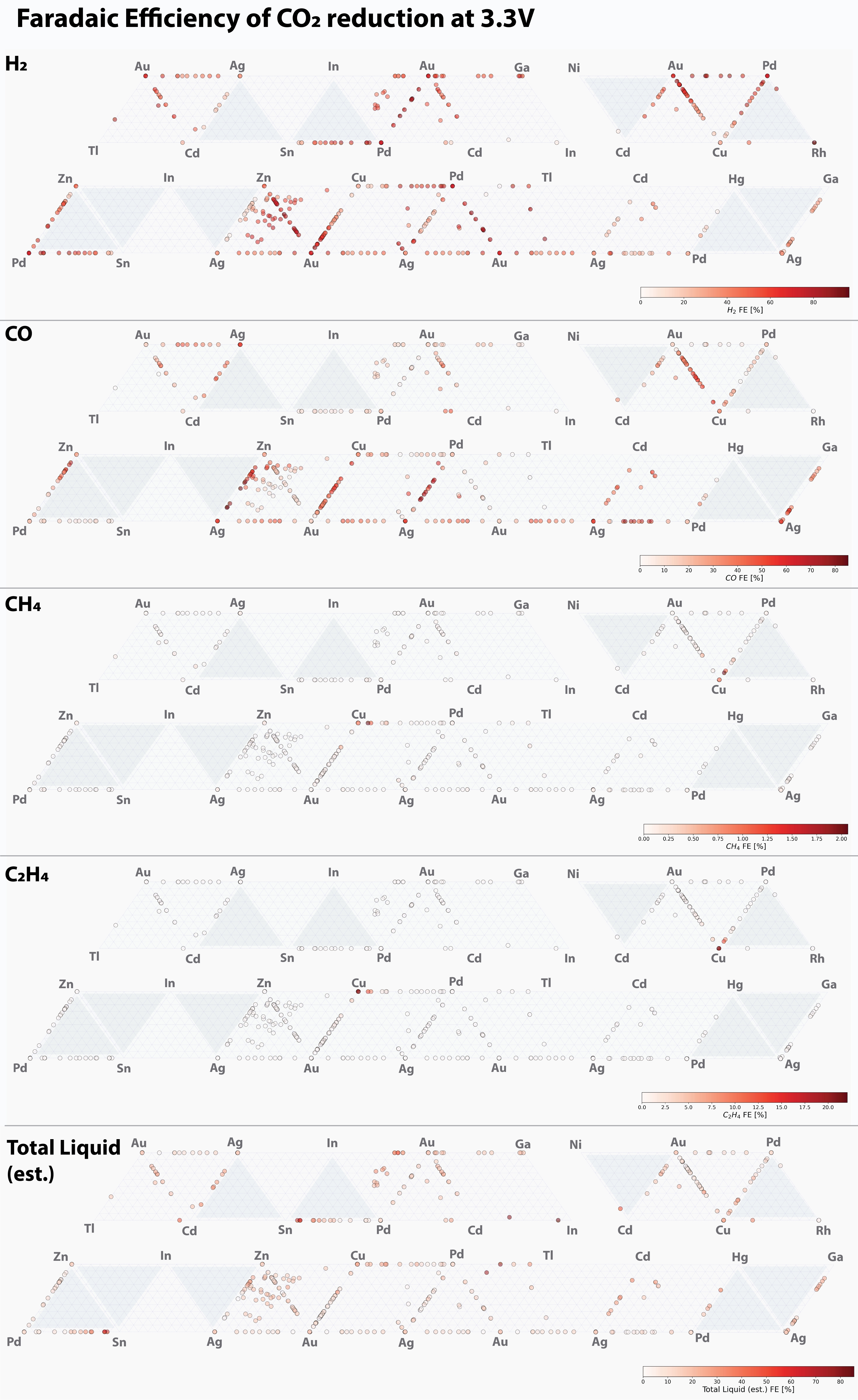}
    \caption{Mapping of Faradaic Efficiency (FE\%) for $H_{2}$, $CO$, $CH_{4}$, $C_{2}H_{4}$, and estimated Total Liquid onto the XRF composition space. Darker
shaded ternary phase diagrams result from connecting several diagrams and were not intentionally explored}
    \label{fig:fe_space}
\end{figure}

To evaluate our predictive models for \cdrr, we first address a few notable experimental differences to \her. While \her~has only one measurable outcome, the experimental voltage, \cdrr~has many possible products including CO, $\mathrm{H_2}$, $\mathrm{CH_4}$, $\mathrm{C_2H_4}$, and liquid products. For a given experimental sample, \cdrr~testing was performed on each sample at several different current densities. To train a single model across the different compositions, we normalize the kinetic driving force, or potential, of the reaction. Given our experimental setup aimed to mirror industrial relevant conditions with two electrodes, fixing the potential through a three-electrode measurement was out of scope. To address this, we interpolate each product to a fixed applied potential of 3.3 V full cell voltage, the average potential across all testing experiments (see Section \ref{sec:SI_fixedpotential} and Figure \ref{fig:fe_space}). This was performed for each sample and the interpolated results were used as targets for the regression models.

For each product, we independently fit a random forest regressor to predict the production rate. We directly measured the quantites of H$_2$, CO, CH$_4$, and C$_2$H$_4$ using gas chromatography. We attempted to fit models for each of these products and also Total Liquid (est.), which is the balance over electrons assuming that the remaining product is formate (two electrons transferred). Boltzmann weighted adsorption energies and Matminer elemental features were used as inputs. The results are shown in Figure~\ref{fig:co2rr-main} for a few products under different cross-validation strategies. When evaluating \gls{LOO} performance, we capture fair correlations for $\mathrm{H_2}$ (R$^2$ = 0.46), CO (R$^2$ = 0.44), and the liquid products (R$^2$ = 0.43). In general, results using spark ablation (circle markers) have stronger correlation than those using chemical reduction (triangle markers). However, results using \gls{LOCO} suggest little correlation for $\mathrm{H_2}$ (R$^2$ = 0.22), and the liquid products (R$^2$ = 0.18) and no correlation for CO (R$^2$ = 0). The results for $\mathrm{C_2H_4}$ are similarly poor, with R$^2$ = 0.08 and 0.01 on \gls{LOO} and \gls{LOCO}, respectively. To predict the noise in our experimental results, we compute the standard deviation from 297 testing replicates within $\pm$5\% in XRF composition. We observe an experimental standard deviation of $\sigma=0.050~(\mathrm{H_2}), 0.040~(\mathrm{CO}), 0.033~(\mathrm{Liquid})$ $\mu$~mol/cm$^{2}$s as shown by the shaded areas in Figure \ref{fig:co2rr-main}.

Rather than fitting on production rates, we also explored fitting directly on the Faradaic efficiencies for each product, with similar results (Figure \ref{fig:fe}). When considering only the elemental features from Matminer, results are comparable to those when considering adsorption energies (Figure \ref{fig:matminer}). This is a good indicator that there exists a large gap between the simple adsorption energies and the reality of \cdrr. Additional results are highlighted in Section \ref{sec:SI_results}; each with similar performance to those highlighted here.

Unlike \her, \cdrr~presents significant obstacles in building experimentally predictive models. These results suggest that leveraging only elemental features and adsorption energies of lone adsorbates on clean surfaces is not sufficient to predict the complexity of this chemistry. Despite these results, we hope they, alongside the experimental dataset, provide an initial baseline for the community to evaluate more sophisticated predictive models.

\subsection{Descriptor Importance}

For both \cdrr~and \her~we compare their performance using different adsorption energy and elemental features to predict $\mathrm{H_2}$ and Voltage, respectively. We focus on $\mathrm{H_2}$ for \cdrr~due to the weak signals for the other products. The importance of features is evaluated for both \gls{LOO} and \gls{LOCO}. A linear model is used for \her~and a random forest model for \cdrr, giving their respective best results. Results are summarized in Table \ref{tab:features}.

For \her~we see fairly consistent results across the board, with $\mathrm{R^2}\sim0.6$ for all \gls{LOO} results, with the exception of the Boltzmann and Wulff weighting ($\mathrm{R^2}=0.26$ and 0.44, respectively). The naive $\mathrm{E_{ads,mean}}$ performs the best on \gls{LOCO} with a $\mathrm{R^2}=0.59$ and \gls{LOO} with the addition of Matminer features of $\mathrm{R^2}=0.63$. \cdrr~results follow different trends. We observe that using only adsorption energy features provides the worst results on both \gls{LOO} and \gls{LOCO}, $\mathrm{R^2}\sim0.3$ and $\mathrm{R^2}\sim0$, respectively. Only with the inclusion of Matminer features does a weak correlation appear with $\mathrm{R^2}\sim0.46$ and $\mathrm{R^2}\sim0.22$ for \gls{LOO} and \gls{LOCO}, respectively. These results suggest that for \her~the proposed adsorption energy descriptors are in fact beneficial and correlate well to experiments. In the case of \cdrr, however, these descriptors have almost no correlation, specifically when looking at \gls{LOCO}. 

The Wulff shape can provide the theoretical, equilibrium nanoparticle structure of a material and the frequency of specific surfaces\cite{tran2016surface, ringe2011wulff}. Unfortunately, we did not see clear improvements with all methodologies when using Wulff or Boltzman weighting over taking the simple mean of the adsorption energies. This shows a need to construct different methods of aggregating surface information to the material property level. We also saw no improvement when considering a single characteristic surface (the closed-packed surface in our case) as is done for small screening studies\cite{schumann2018selectivity, vojvodic2014exploring, liu2017understanding} which develop activity or selectivity projections as a function of 1-2 descriptors. 

%%%%% Conclusions
\section{Conclusion}

In this work, we strive to bridge the gap between computational models and experimental studies. While datasets for building computational ML models have grown substantially\cite{chanussot2021open,omat24,spice, smith2020ani,trans1x}, constructing larger datasets of experimental results has proven challenging. This severely limits our ability to train models that map computational descriptors to experimental results. To address this, we perform a high-throughput experimental study at industrially relevant conditions for \her~and \cdrr~covering 441 catalyst samples. Each sample was tested in a consistent manner and characterized by XRD and XRF to facilitate their connections to computational descriptors. In addition to a large experimental dataset, we present the largest computational catalyst screening campaign to date. It leverages machine learning to consider a previously intractable material design space of 19,406 materials. For each material, we calculated the adsorption energies of six adsorbates, including a multicarbon adsorbate, on all surfaces up to Miller index 2 with DFT validation. 

The simulations for this study neglect several intricacies, which have been shown to have notable impacts on a catalyst's activity. The electrolyte, pH, gas-solid-liquid interface, temperature, and reactor design all have an impact on performance\cite{nitopi2019progress, billy2017experimental}. These findings extend to the computational realm as well. Theoretical work has shown that electric field effects at the surface stabilize important intermediates\cite{sandberg2016co, chen2016electric}. In a related manner, the pH and electrolyte have also been shown to impact product selectivity and activity\cite{liu2019ph, xu2022theories}. Lastly, inclusion of solvent in simulations and the solvent approach can have a large impact on the resultant selectivity and/or activity\cite{dyson2016solvent, saleheen2018liquid}. 

Despite our simplifications, we were able to develop a strong correlation between \her~experimental results and computational adsorption energies (H and OH). Using the correlation, we were able to predict the experimental activity of the 19,406 materials considered in the computational pipeline. This reveals a Sabatier volcano on which Pt is a top catalyst, despite the fact that Pt and Pt-alloys are absent from the training data. A list of materials are provided that are predicted to be more active than platinum. 

Correlations for \cdrr~using adsorption energies from six adsorbates are weaker and demonstrate the challenges and opportunities for future research in predicting the selectivity of complex reactions. In addition to the complexity of electrochemical conditions, \cdrr~is difficult to treat because of its reaction network complexity. Producing a product selectively is a grand challenge because there are many competing possible products, chief among them being hydrogen gas\cite{nitopi2019progress}. Even when only considering products with up to two carbons, there are hundreds of possible reactions to make those products. There is no clear consensus on which reaction pathways are important for just pure copper\cite{nitopi2019progress, peng2021role, deng2024unraveling}, and even less clarity on which pathways are important for an arbitrary material.

There is opportunity to improve the work presented here, in an effort to realize AI-driven discovery for \gls{CO2RR}, by constructing a larger experimental dataset. As shown for \her~in Figure \ref{fig:scaling}, the addition of more data results in improved correlations. More data would also allow the possibility of using more complex modeling techniques that may better fit \cdrr~data. In addition, the impact of neglecting electrochemical conditions in the computational pipeline should be considered. Perhaps considering the network explicitly would improve our ability to make connections. This could be tractable since \gls{MLP} can be used to rapidly access transition state information\cite{wander2024cattsunami}. There is also opportunity to explore different modeling techniques and in particular how to perform the abstraction from computational surface-level properties to aggregated material-level production rates. This problem is difficult, but important. Hopefully the creation of this dataset and datasets like it in the future will allow computationally driven experimental discovery to become a reality.

% \clearpage

\bibliographystyle{unsrt}
\bibliography{bib}

\begin{thebibliography}{100}
\bibitem{vasp-license}
``{The} calculations in this work have been performed using the ab-initio total-energy and molecular- dynamics package {VASP (Vienna ab-initio simulation package) developed at the Institut f{\"u}r Materialphysik of the Universi{\"a}t Wien2,3."}.
\end{thebibliography}

% \clearpage
\newpage
\beginappendix
\section{Experimental Studies}

\subsection{Experimental Material Selection} \label{sec:SI_material-selection}

Diversity is essential when creating training data for ML models. In our experimental studies, this includes the diversity of the elemental composition and the diversity of the products produced when they are used as catalysts. To increase the likelihood of sampling materials that produce different products, we grouped the materials into three rough groups and attempted to sample them equally. For \cdrr, we devised a multi-objective optimization framework to suggest materials that are likely to produce multi-carbon products. Through this analysis, we selected 317 compositions as possible C$_{2+}$ candidates based on the following scoring function:
\begin{align}\label{eq:score}
    score = w_1||E_{CO} + 0.67||_2 + w_2||E_{H} - 0.15||_2 + w_3(E_{COCOH} - 2E_{CO})
\end{align}

Here, $E_x$ corresponds to the computationally calculated adsorption energy of a particular adsorbate. The three terms in Equation (\ref{eq:score}) correspond to (1) targeting a CO adsorption energy near -0.67 eV, (2) targeting a H adsorption energy near 0.15 eV, and (3) favoring COCOH binding affinity over CO. Each term is normalized to ensure that similar distributions do not bias the overall objective function and objective weights $w_i$ are used to place more importance on descriptors. Initially, these weights are set to $w_i=1$, but could have been updated in response to experimental findings. Once all surfaces have been assigned a score, we classify the top 10\% of the materials as C$_{2+}$ candidates.

Missing from the expected outcome is the presence of many C$_{2+}$ classified Cu alloys. There are only five such alloys out of the 317 selected. This could be a reflection of some failing in the framework, but it was intentionally not optimized with this in mind. There is literature precedent for certain Cu alloys experimentally showing a propensity towards multi-carbon products, but often those materials are not ordered intermetallics\cite{hoang2018nanoporous, jeon2022selective, ren2016tuning, zhu2019low, ma2017electroreduction} like those considered here. There are also a modest number of materials containing chalcogens that have been classified as possible C$_{2+}$ producers. There is literature precedent for Se and Te alloys producing multi-carbon products\cite{saxena2021selective, saxena2022nickel, saxena2023copper}, but they have also been shown to produce other C$_1$ products\cite{yang2019selective, saxena2022cobalt}. The product selectivity for this family of materials appears to be very sensitive to exact composition ratios and reaction conditions. The prospect of discovering multicarbon selective alloys composed of earth-abundant elements such as Ga and Zn is exciting, but alloys studied containing these elements have been shown to suffer from poor Faradaic efficiencies\cite{he2018electrocatalytic}. Nickel gallium alloys have been shown to produce multi-carbon products\cite{torelli2016nickel} with very low onset potentials, but they suffer from poor selectivity\cite{van2024experimental}. Without extensive experimental testing, it is unclear whether the materials suggested by the multi-objective framework to be possible C$_{2+}$ producers will be realized as such.

For \her~we categorized all compositions with more than 50\% of surfaces having H adsorption energies less than -0.1 eV as likely hydrogen producers during \cdrr. All other compositions were classified as possible C$_1$ producers. This resulted in 11,916  \her~compositions and 3,201 C$_1$ compositions. Note, these sets of compositions only provide potential candidates for synthesis of which only a subset were selected. At a high level, chemical intuition is retained. For \her, there are a large number of Pt and Pd alloys. This is sensible because unary Pt and Pd are known hydrogen producers\cite{hori1994electrocatalytic, kuhl2014electrocatalytic}. For C$_1$, there are a large number of gold and silver alloys, which as unaries are known to produce carbon monoxide\cite{hori1994electrocatalytic, kuhl2014electrocatalytic, thevenon2021dramatic}. For C$_1$ there is a general shift toward less reactive elements compared to elements that appear frequently for \her. This shift continues for C$_{2+}$ products, which tend more toward post-transition metals and metalloids.

Our goal was to roughly select an equal number of materials from each category (C$_{2+}$, C$_1$, and HER). However, the C$_{2+}$ samples proved difficult to synthesize. The accessibility of elements for both synthesis techniques is shown at the top of the Experimental Chemical Space section of Figure \ref{fig:chemical_space}. For spark ablation, there were substantial oxidation issues when synthesizing Zn, In, and Sn. Gallium was also difficult to incorporate because it has a low melting temperature. These four elements were present in nearly all of the materials that we classified as possible multi-carbon producers. For chemical reduction, there were similar challenges preparing samples that were classified as possible multi-carbon and C$_1$. Note, while Pt was accessible to our synthesis techniques, no samples were synthesized using Pt in this study.

Our synthesis techniques are limited in their ability to match target compositions with errors up to $10\%-15\%$. Due to this, additional samples were synthesized in the "gaps" between target compositions. This increased the likelihood that samples would be synthesized close to the target compositions, as verified by \gls{XRF}. A byproduct of this approach is that a considerable number of synthesized samples are not in the three categories above (C$_1$, C$_{2+}$, and \gls{HER}).

\subsection{Spark Ablation Synthesis} \label{sec:SI_sparksynth}

\subsubsection{Printing of nanoporous films}

In this work, the VSP-P1 Nanoprinter was used to synthesize inorganic nanoporous thin films with different binary and ternary elemental composition. The P1 uses three sequential automated steps to convert a solid rod into particles smaller than 5 nm and print them in the form of nanoporous layers.

In the first step, two solid rods are placed with the tips closely spaced to each other. Electrical sparks are formed between the rods to create a local plasma. Due to the high temperature of the plasma (>20,000$\degree$C), small parts of the rods are evaporated into an atomic cluster.  Directly after the spark, the loose atoms bump into each other and grow into bigger clusters/nanoparticles.

In the second step, the evaporated material is transported by a carrier gas that flows in between the electrodes. Since the carrier gas is at room temperature, it accelerates the cooling of the hot gas after the plasma stage. Due to this fast quenching, the particles of the cluster stabilize (e.g. stop aggregating)  at 2 – 5 nanometers in size. During their time-of-flight, these stable primary nanoparticles could still form agglomerates by sticking to and colliding with each other.

In the final step, the impaction method is used to get the growing particles out of the carrier gas and printed on a substrate. This is achieved by accelerating the gas flow to supersonic speeds by forcing them through a nozzle with a narrow orifice. At this speed, the growing nanoparticles are not able to follow the gas flow but instead follow a straight trajectory and immobilize on the substrate. In this way, thin-film nanoporous layers of nanoparticles can be formed with different thicknesses. Since the substrate is mounted on an XYZ-gantry stage, the nanoparticles can be printed in all kind of complex patterns. It is important to mention that because of the small size of the nanoparticles, they have a very strong Van der Waals force which make them stick to any surface or other nanoparticles. This unique feature enables the VSP-P1 to print sticking layers on almost any substrate without adding binders.
 
A dedicated P1 was used for this project with the following specs:
\begin{itemize}
\item 3 units of VSP-G1 nanoparticle sources
\item Carrier gas: Ar/H2 (95/5)
\item Nozzle diameter: 0.55mm
\item Flow rate in G1: 1.35 L/m
\item Substrate to nozzle distance: 2.4 mm
\item Current : 0.1 – 10 mA
\item Voltage: 0.5 – 1.3 kV
\item Vacuum chamber pressure: 1.65 - 1.70 mbar
\end{itemize}
(Note: Voltage and current are adjusted according to the desired output to achieve specific loading and alloying ratios)

\subsubsection{Calibrating the printer for different materials}

Due to differences in evaporation enthalpy, it is important to calibrate the amount of material that is deposited on the substrate. This calibration is specific for element type, electrode diameter, carrier gas type, flow, nozzle size, and distance between nozzle and substrate.
Calibration is performed by printing on carbon paper \gls{GDL} (type H23C3 from fuelcellstore.com) for different periods of time and measuring the mass increase. Plotting these points on a graph with an intercept of zero gives a slope that represents the printing rate in milligrams per hour.
The calibration is done with two of the VSP-G1 sources connected, including multiple times with cleaning of the G1 in between prints. Calibration was performed under fixed operating conditions: the gas flow rate in both G1s were 1.35 L/min Ar/H2 (95/5), and the spark settings were 1.3 kV and 10 mA with a spark frequency of 220 Hz.
 
\subsubsection{Controlling oxidation of the samples during shipment for electrochemical testing}
Since most of the elemental analysis was conducted at the University of Toronto's facilities, we investigated the impact of shipping on the samples. The nanoporous metallic samples were expected to oxidize when exposed to air (oxygen). However, we discovered that packaging the samples under inert gases like Argon still resulted in oxidation, despite using cycles of vacuuming and purging with inert gas. Surprisingly, packaging the samples in air did not lead to oxidation. XRD analysis revealed that the samples remained metallic for more than six days when exposed to air. Annealing the samples under a gas mixture of 95\% Ar and 5\% $H_2$ at 400$\degree$C for 30 minutes after shipment was found to reduce surface oxides and alloy the metal mixtures\cite{dai2023sub}. During the annealing process, the particle size is increased gradually until a layer is formed. By optimizing the annealing temperature the alloy can be formed without losing to much of the porosity. Annealing was limited to relatively low temperatures and for short times to prevent nanoparticle growth and reactivity reduction.

\subsubsection{Controlling the elemental composition by mixing 2 or 3 sources}

A standard VSP-P1 is able to work with 2 VSP-G1 nanoparticle sources. The output of the 2 sources is combined into a single gas stream that is entering the printer nozzle.
If one sources is setup with gold electrodes and the other with copper electrodes the P1 is printing nanoporous layers made of stacked gold and copper nanoparticles.
For this study, VSP modified the P1 to accommodate a 3rd VSP-G1 source to make ternary compositions as well as binary.
The composition of the particles in the gas stream can be tuned by lowering the power settings for one, or two in the case where three sources are used, of the G1s. One G1 is kept at 1.3 kV and 10 mA as max settings. Lowering the output is done by lowering the current and keep the voltage constant. In this way the gap between the electrodes stays constant and only lowers the frequency of sparking. The frequency cannot be set in the G1, but rather it is a result of the combination between voltage and current that is chosen. There is a limit to this, however, if the frequency drops below around 20 Hz the sparking becomes unstable. This happens when the current is below 1.1 mA and the voltage has been lowered to 0.5 kV.  The loading/layer thickness is tuned by the movement speed of the stage during printing.

\subsubsection{Printing the same elemental composition on wafer and GDE substrates}

The initial goal of the synthesis was to deposit sample on a silicon wafer to enable characterization by XRF and XRD without peak interference in the region of interest. This involved printing 2x2 mm squares of various compositions on the wafer, achieving a loading of 0.2 $mg/cm^{2}$ by stacking five layers sequentially. Each batch included 25 unique compositions. After annealing the Si wafer, XRD was used to identify the crystal structure of the samples, which were then replicated on \gls{GDE} for performance testing. However, XRF revealed source output deviations of 1-15\% from the setpoint, complicating the replication of exact compositions on GDEs. 
To resolve this, we began printing wafer and GDE samples simultaneously. Each batch included a 4-inch wafer and a 10x10 cm GDE substrate, allowing consistent and stable G1 settings. Layers were alternated between the wafer and GDE to average out compositional drift.

\subsection{Chemical Reduction Synthesis} \label{sec:SI_chemreduction}

\subsubsection{Automated catalyst preparation: wet-chemistry synthesis}

\begin{figure}
    \centering
    \includegraphics[width=0.8\linewidth]{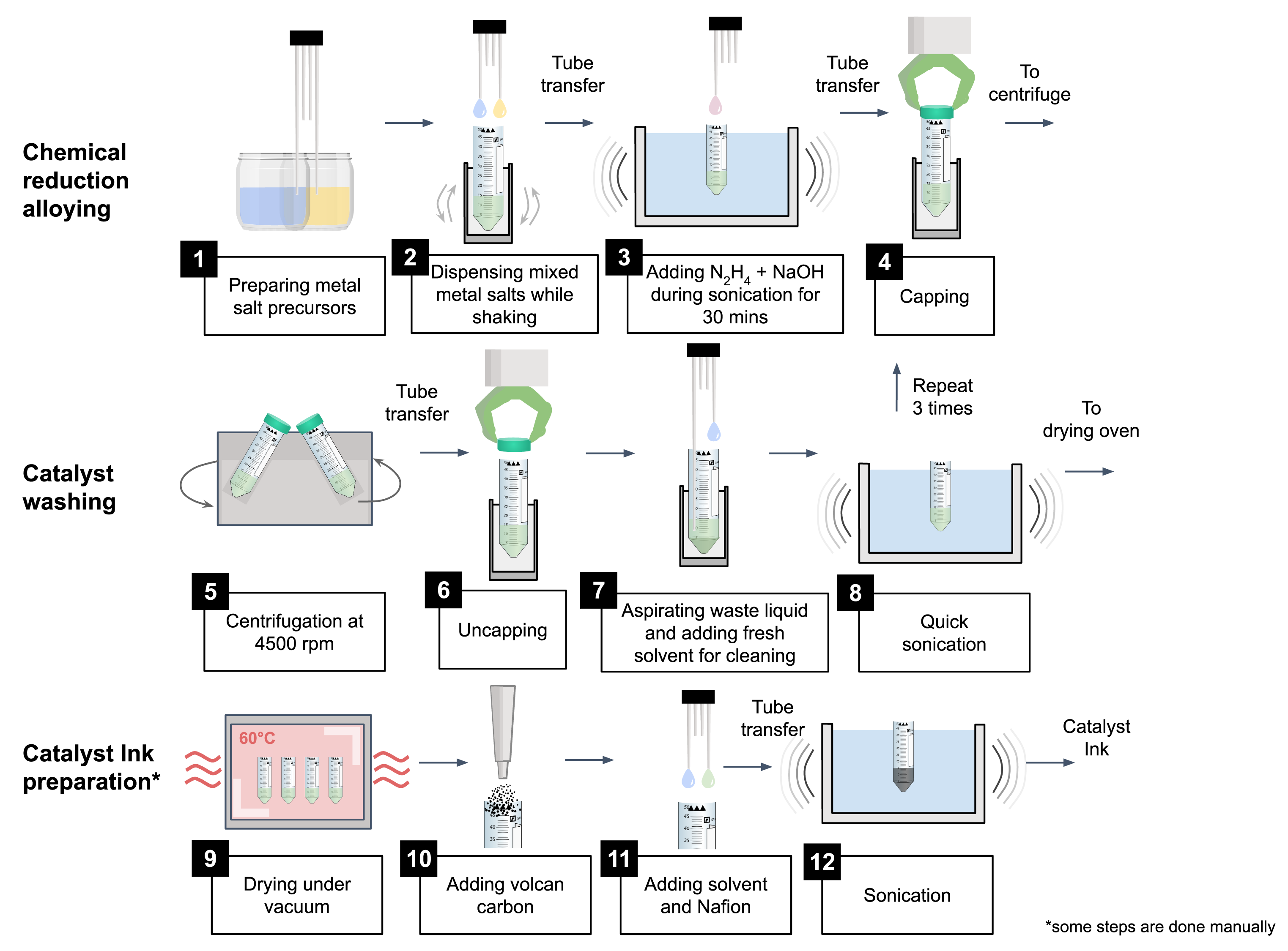}
    \caption{This workflow diagram illustrates the step-by-step process of the chemical reduction synthesis method using the Chemspeed robot.}
    \label{fig:chemical_reduction_workflow}
\end{figure}

\begin{figure}
    \centering
    \includegraphics[width=0.8\linewidth]{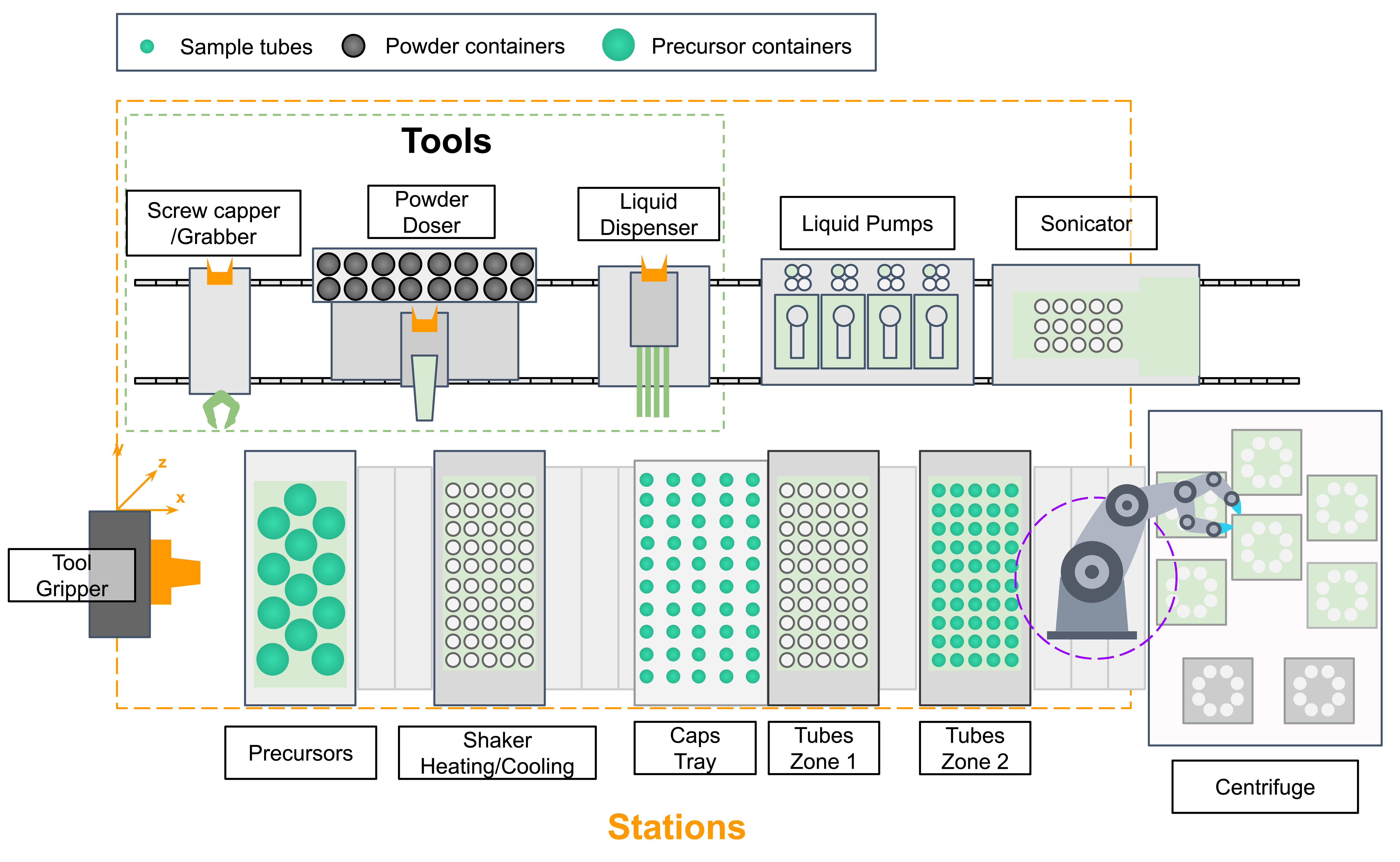}
    \caption{This is a top layout of the Chemspeed robot used for the chemical reduction synthesis.}
    \label{fig:chemspeed_layout}
\end{figure}

\begin{figure}
    \centering
    \includegraphics[width=0.8\linewidth]{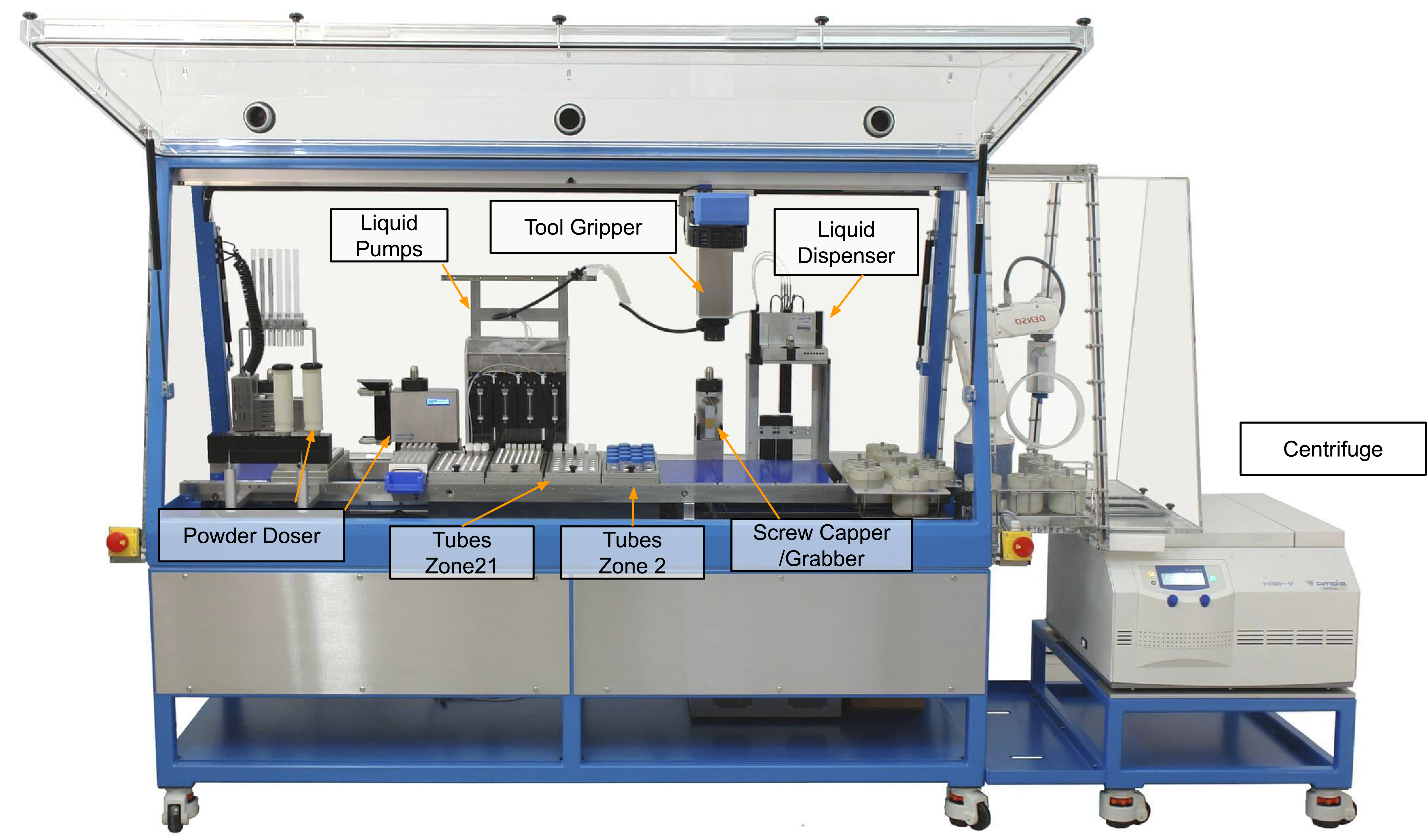}
    \caption{This is a photo of the Chemspeed robot used for the chemical reduction synthesis.}
    \label{fig:chemspeed_photo}
\end{figure}

The automated synthesis was performed using the Chemspeed Swing XL platform (Figures \ref{fig:chemspeed_layout} and \ref{fig:chemspeed_photo}). Metal precursor stock solutions (0.15 M) and a hydrazine/NaOH(2.67 M) reducing solution (25:27, v/v) were manually prepared and loaded into the platform. Precise ratios of these stock solutions were mixed to achieve the desired compositions (1.5 mL total per tube). After thorough mixing, 0.65 mL of hydrazine/NaOH solution were added and ultrasonicated at 50$\degree$C for one hour. The process concluded with three purification cycles involving ultrasonication and \gls{DI} water rinses and, lastly, the solutions were dried in a vacuum oven (Figure \ref{fig:chemical_reduction_workflow}).

\subsubsection{Catalyst annealing}

The catalysts prepared via both synthesis approaches were annealed under a mixed gas atmosphere of 5\% H2 and 95\% Ar (forming gas) to induce the formation of alloys. For spark ablation, samples were annealed after being printed on a Si wafer and \gls{GDL}. For samples prepared via chemical reduction, the resultant powder was annealed prior to drop-casting on a \gls{GDL}. Initially, the tube furnace was purged with the forming gas for 3 hours and then the temperature was ramped to 400$\degree$C at a rate of 5$\degree$C/min, followed by aging at 400$\degree$C for 30 minutes. After the heat treatment, the furnace was cooled down to room temperature while continuously purging with forming gas to prevent oxidation of the samples before they were removed from the furnace.

\subsubsection{Catalyst ink and electrode Preparation}

For the catalysts obtained via chemical reduction synthesis, a catalyst ink was prepared by mixing the synthesized powde with  Vulcan carbon XC-72R, Nafion perfluorinated resin solution (5 wt\%), and 2-propanol in a ratio of 6 mg : 4 mg : 20 uL : 500 uL. The mixture was then ultrasonicated for 1 hour and drop-casted onto a 1 cm$^2$ \gls{GDL} (Freudenberg H23C3).

\subsection{Electrochemical Testing} \label{sec:SI_testing}
As an improvement to the previously reported testing apparatus\cite{kose2022high}, gas vials are directly connected to the \gls{MEA} cells, enabling simultaneous sample collection and streamlining the process. This allows for asynchronous testing by storing samples between runs, ensuring continuous operation without manual transfers to a \gls{GC}. The vials can be unplugged and moved to an autosampler for sequential injection into the \gls{GC}, enhancing testing throughput and supporting full parallel operation and asynchronous operation.

\section{Characterization} \label{sec:SI_characterization}
\subsection{XRF and XRD experimental setup}
X-ray fluorescence spectroscopy (\gls{XRF}) was conducted using a Fischerscope X-ray XDAL with a microfocus tungsten (W) X-ray tube operating at 50 kV. Each sample was measured three times with an acquisition time of 20 seconds per measurement, while scanning the stage in xy directions. X-ray diffraction (\gls{XRD}) measurements were performed using a Bruker D8 Discover diffractometer equipped with a Cu microfocus X-ray source ($\lambda$ = 0.15418 nm), operating at 50 kV and 1000 $\mu$A. A 0.4 mm slit was employed, and data were collected over a 2$\theta$ range of 5$\degree$ to 80$\degree$ with a step size of 0.04$\degree$ and a dwell time of 1.2 seconds per increment. The diffraction patterns were recorded at ambient temperature in 1D mode, with the detector aperture set to 62 mm x 20 mm. The Si wafer with multiple catalyst samples was positioned on the measuring stage and the coordinates of five initial samples were determined using the overhead camera built into the diffractometer. A script was used to apply a translation and 2D rotation to calculate the coordinates of the remaining samples.

\subsection{XRD analysis}

\begin{figure}[h!]
    \centering
    \includegraphics[width=\linewidth]{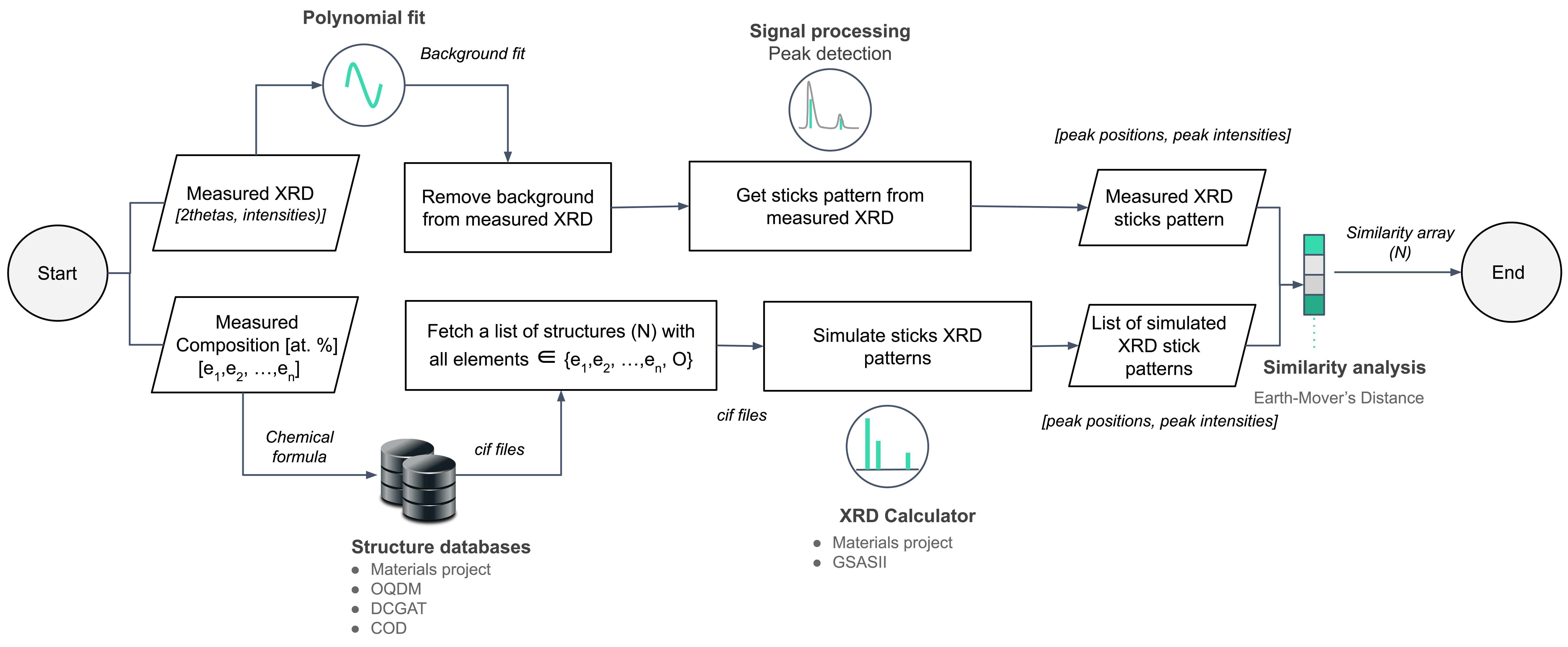}
    \caption{A flowchart of the first step of the \gls{XRD} analysis pipeline. This steps constructs a similarity analysis using the experimental \gls{XRD} pattern and a list of simulated \gls{XRD} patterns from structures that are fetched from Materials Project, OQMD, and Alexandria}
    \label{fig:xrd_step1}
\end{figure}

\begin{figure}[h!]
    \centering
    \includegraphics[width=0.8\linewidth]{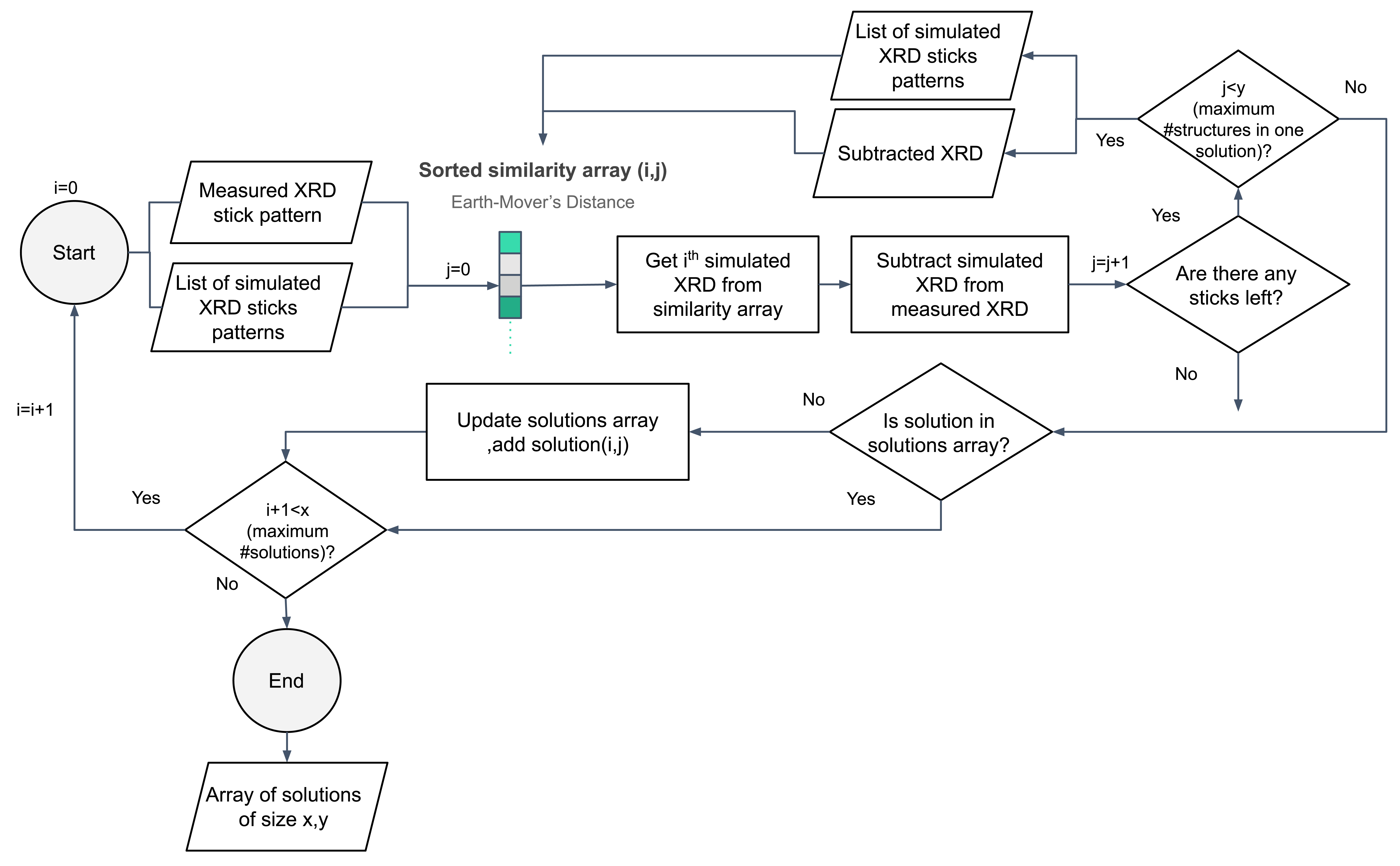}
    \caption{A flowchart showing the iterative process of matching the experimental \gls{XRD} pattern with multiphases. The process involves iteratively removing matched patterns, updating the experimental pattern, and identifying new matches to add new matched phases.}
    \label{fig:xrd_step2}
\end{figure}

\begin{figure}[h!]
    \centering
    \includegraphics[width=0.8\linewidth]{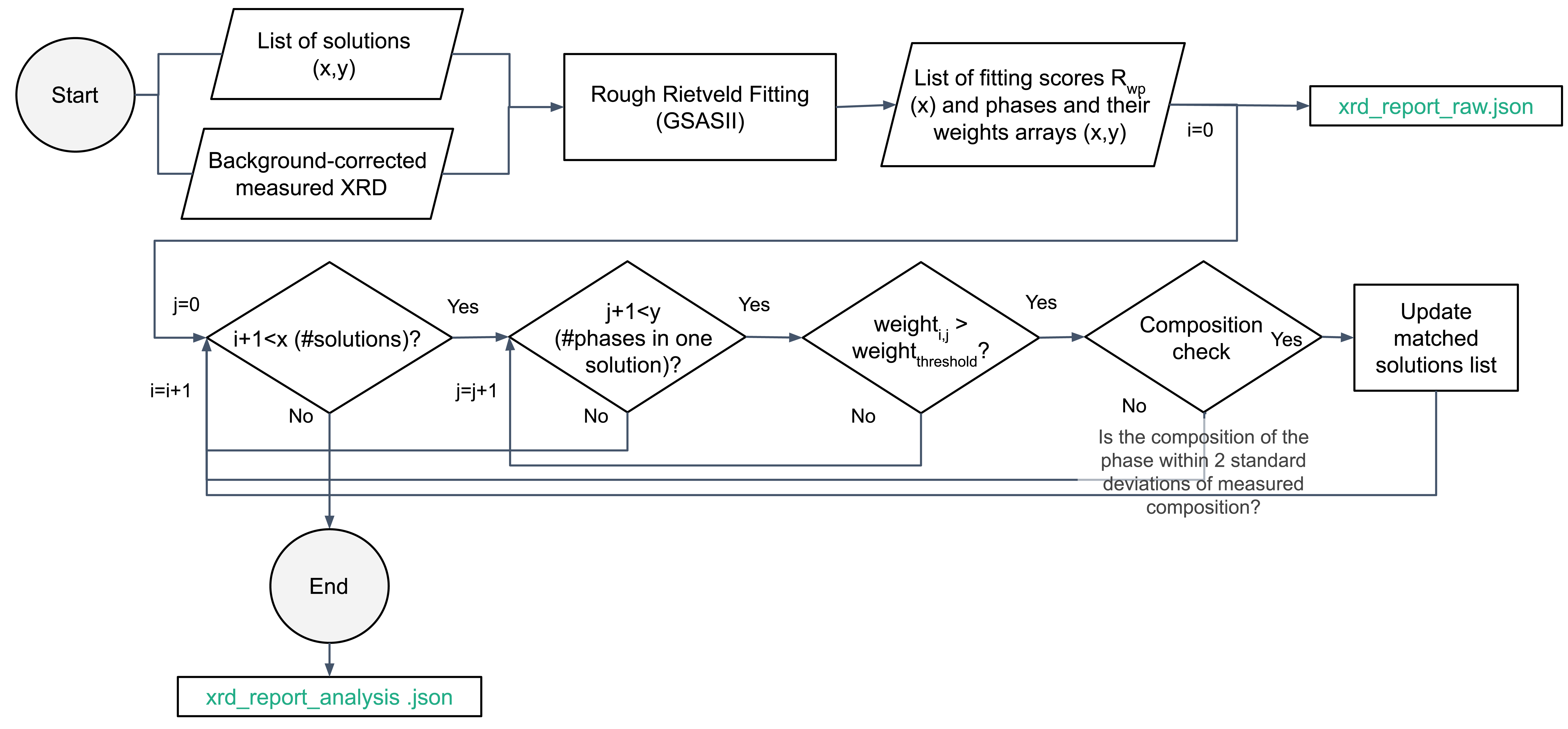}
    \caption{A flowchart showing the final step in the \gls{XRD} analysis running a Rough Rietveld Refinement to evaluate the goodness of the fit and aid with ranking solutions.}
    \label{fig:xrd_step3}
\end{figure}

The process begins by cleaning the \gls{XRD} data to remove background noise and extract clear diffraction patterns. We then fetch computational structures that contain the elements of interest and simulate their \gls{XRD} patterns using a tool in pymatgen \cite{pymatgen}. We use Earth-Mover's distance to calculate how similar these patterns are to the experimental data \ref{fig:xrd_step1}.
After identifying a phase, we remove overlapping peaks and repeat the first step until all phases are matched or a set limit is reached \ref{fig:xrd_step2}. Rietveld fitting helps us assess the quality of the match by calculating the weighted profile R score ($R_{wp}$), which evaluates the quality of the fit \ref{fig:xrd_step3}. This entire process is repeated for a user-defined number of trials to run through a set number of possible phase combinations without excessive computational cost.

\section{Computational Studies}
\begin{figure}[h]
    \centering
    \includegraphics[width=0.6\linewidth]{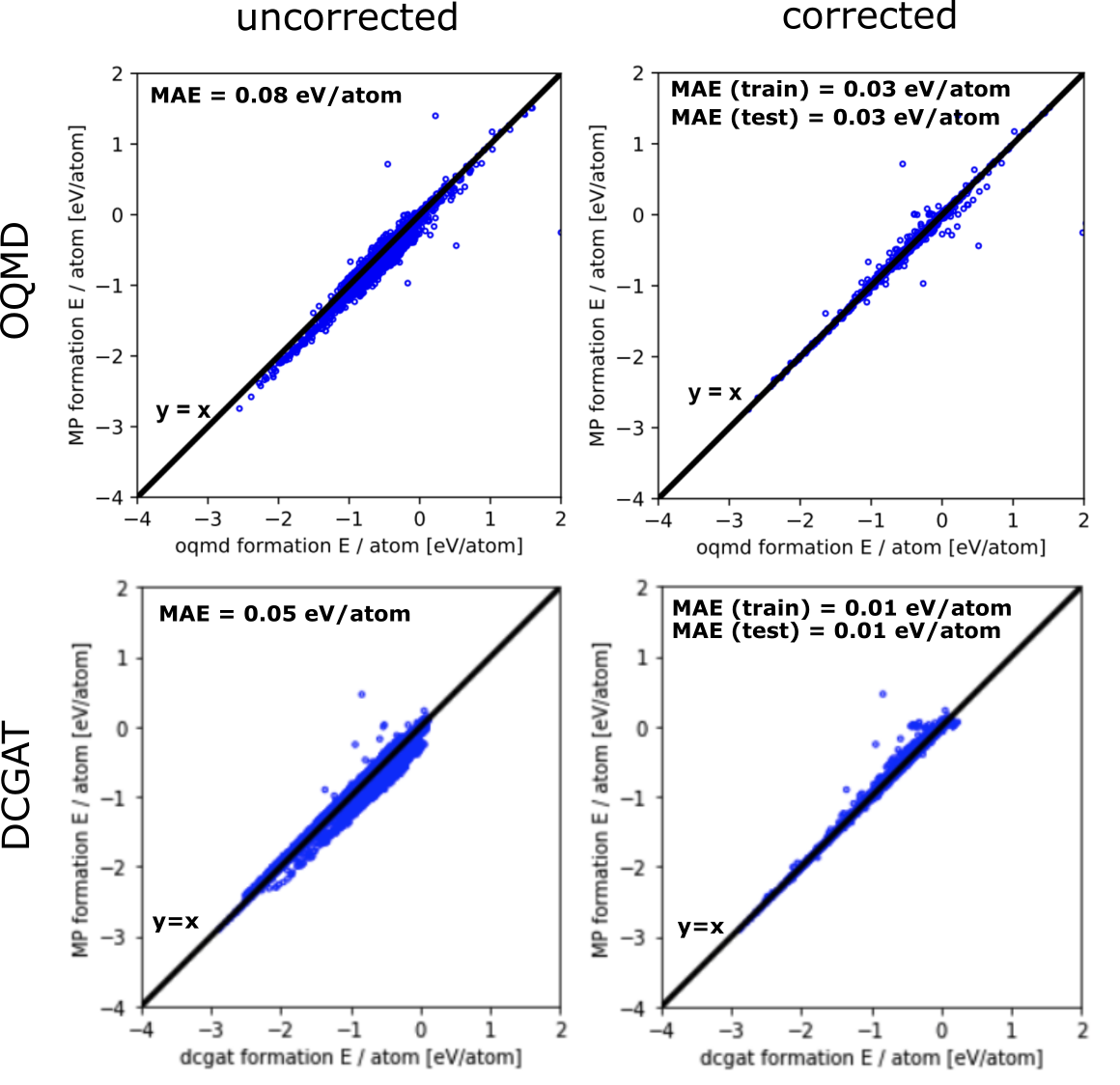}
    \caption{The linear formation energy corrections applied to the materials coming from \gls{OQMD} and Alexandria (DCGAT) to correct small differences in the pseudopotentials so that comparisons could be made to \gls{MP} materials for the purposes of assessing Pourbaix stability.}
    \label{fig:form_e_corrections}
\end{figure}
\subsection{Material Selection} \label{sec:compmatselection}
We considered all materials available in three permissively-licensed computational materials databases: \gls{MP}~\cite{MP_main}, \gls{OQMD}~\cite{oqmd1, oqmd2}, and Alexandria~\cite{dcgat, cgat}. In an effort to improve the likelihood of discovering an industrially viable catalyst, materials were filtered by assessing their thermodynamic stability under reaction conditions. To do this, we evaluated the decomposition energy using the Pourbaix framework implemented in pymatgen\cite{pbx_efficient, pymatgen, pbx_main}. We were interested in neutral pH conditions (in the bulk), but the pH will be elevated at the electrode under operating conditions\cite{billy2017experimental, zhang2020direct}. Materials with a decomposition energy less than 0.05 eV/atom anywhere in the operating range of pH (7 to 14) and applied potential (-1.35 to -0.55 V versus SHE) were retained. To facilitate this analysis despite any small differences in \gls{DFT} settings, we determined a subset of materials within the non-MP databases with analogous entries in MP. Using the formation energies of these materials in each of the databases, we fit linear corrections to the formation energies to be MP-like (see Figure \ref{fig:form_e_corrections}). From the Pourbaix analysis, we found 2,458 materials from \gls{MP}, 3,173 materials from \gls{OQMD}, and 13,775 materials from Alexandria for use in our computational pipeline. The bulk materials were then re-optimized with \gls{RPBE} to be consistent with the level of theory of \gls{OC20}, which the \gls{ML} models used in this work were trained on. We removed materials from Alexandria that have magnetic moments because of an error in calculation. For MP, OQMD, and the data used to train the models used for structure optimizations magnetic moments were neglected. For the materials with Alexandria, the bulk optimizations were performed with magnetic moments, an inconsistency with the rest of the pipeline. Still, for better results on magnetic materials they should be handled more appropriately.

\subsection{Adsorption Energy Calculations} \label{sec:adsorbcalc}
\subsubsection{Adsorbate placements}
Catalyst slabs were created from relaxed bulks using pymatgen\cite{pymatgen} implemented code to enumerate all surfaces up to Miller index 2. All surface terminations were used for materials from \gls{MP} and \gls{OQMD}. For materials from Alexandria, two surface terminations per Miller index were selected: a random termination and the lowest surface energy termination. All surface and placement enumeration code is provided at https://github.com/Open-Catalyst-Project/Open-Catalyst-Dataset. Materials coming from MP and OQMD materials were done from an earlier snapshot of the codebase (commit 17e350e). Alexandria enumeration was performed after a refactor (commit b12aac4), which provided improved random placements and minor fixes throughout. A total of $\sim700,000$ unique surfaces were generated across the selected materials. 

Placements were made with several modes: full heuristic placements, random sites with heuristic placements, and full random placements. For random sites, a Delaunay meshgrid is constructed with sites uniformly sampled across the grid. For heuristic, sites are assigned at each surface atom (atop), between every pair of adjacent surface atoms (bridge), and at the center of every group of three or four adjacent surface atoms (hollow). For all approaches, the adsorbate binding atom is placed onto the site, then the adsorbate is translated along the surface normal until the minimum distance between any adsorbate atom and any surface atom is 0.1 \AA{}. Additionally, to increase the diversity of initial configurations considered, the adsorbate is uniformly randomly rotated around the z-direction and provided a slight wobble around x and y, which amounts to a randomized tilt within a certain cone around the north pole. For even greater diversity, random rotations in all directions about the center of mass are applied for the fully random placement mode. Placements made before updates followed a similar strategy but used CatKit\cite{catkit} for heuristic placements and restricted random rotations to only the z-axis. These changes were shown to have negligible effect on the performance of adsorption energy calculations, as shown by Lan et al.\cite{lan2023adsorbml}.

\subsubsection{ML relaxations}
Screening the proposed search space for \gls{CO2RR} with \gls{DFT} would be an intractable problem for even the world's most powerful supercomputers. To address this, recent work \cite{chanussot2021open, lan2023adsorbml} has demonstrated the utility of generalized \gls{ML} potentials for catalyst discovery. The incredible progress the community has made in the development of more accurate models has made the reality of large screening campaigns like this possible.

For all adsorbate and surface combinations, we calculate the adsorption energies using an AdsorbML\cite{lan2023adsorbml} ML+SP approach. Given an adsorbate-surface combination, we first enumerate $\sim$165 different adsorbate configurations (100 random + $\sim$65 heuristic) on the surface. \gls{ML} relaxations are then performed using GemNet-OC\cite{gasteiger2022gemnet}, a cost-effective \gls{GNN} trained on the \gls{OC20} dataset. Unlike a \gls{DFT} relaxation that can take roughly 24 hours per relaxation, \gls{ML} models used in this work finish in as little as five seconds. Once the \gls{ML} relaxed states are computed, the five lowest-energy systems are selected and a single point \gls{DFT} call is made on the \gls{ML} relaxed structure to get a more accurate adsorption energy. For an in-depth discussion of the algorithm and model, we refer the readers to the original AdsorbML\cite{lan2023adsorbml} and GemNet-OC\cite{gasteiger2022gemnet} manuscripts. All \gls{ML} relaxations were run for 200 optimization steps or until a maximum force norm of 0.02 eV/\AA{} is achieved, whichever comes first.
%%% Discuss AdsorbML and small details

\subsubsection{DFT calculations}

\gls{DFT} relaxations were performed consistent with \gls{OC20} and AdsorbML. \acrfull{VASP} with \gls{PAW} pseudopotentials and the \gls{RPBE} functional were used for all calculations~\cite{Kresse1994, Kresse1996a, vasp-license, kresse1999ultrasoft, Kresse1996}. All single-point calculations were performed with a maximum number of electronic steps of 300 to ensure that the initialized wavefunction had sufficient steps to converge. Single-point calculations with unconverged electronic steps were discarded.

% \begin{enumerate}
%     \item Failed jobs $-->$ run custodian
%     \item Some jobs had KPOINTS coarser than desired (rerun?)
% \end{enumerate}

\subsection{Interpolation to fixed applied potential} \label{sec:SI_fixedpotential}
All experiments were performed at fixed potentials of 50, 100, 150, 200, and 300 mA/cm$^2$. To make comparisons with a fixed driving force, a linear interpolation to a fixed potential was applied. A fixed potential of 3.3 V was used, the average potential across all testing experiments. This interpolation was limited so extrapolation would not be performed and was performed in log space, so the relation between voltage and current density should be linear. If the chosen potential occurred at a current density outside of the tested range (50-300 mA/cm$^2$), then the value at the closest end of the range was taken. This was performed for each sample and the interpolated results were used as targets for the regression models.

\subsection{Additional results}\label{sec:SI_results}
The full set of results for both \cdrr~and \her~across both cross-validation strategies are summarized in Table \ref{tab:full-results}. Results on $\mathrm{CH_4}$ and $\mathrm{C_2H_4}$ can appear positive but are primarily a result of little to no production, allowing the models to predict near 0 successfully. A larger dataset could remedy this by providing more samples for these products.

% \usepackage{multirow}
% \usepackage{graphicx}
\begin{table}[]
\centering
\resizebox{0.6\textwidth}{!}{%
\begin{tabular}{lcccccc}
 &
   &
   &
  \multicolumn{2}{c}{\gls{LOO}} &
  \multicolumn{2}{c}{\gls{LOCO}} \\ \hline
\multicolumn{1}{l|}{Source} &
  \multicolumn{1}{c|}{Reaction} &
  \multicolumn{1}{c|}{Product} &
  $\mathrm{R^2}$ &
  \multicolumn{1}{c|}{MAE} &
  $\mathrm{R^2}$ &
  MAE \\ \hline
\multicolumn{1}{l|}{\multirow{9}{*}{Spark + Reduction}} &
  \multicolumn{1}{c|}{HER} &
  \multicolumn{1}{c|}{Voltage [V vs SHE]} &
  0.61 &
  \multicolumn{1}{c|}{0.08} &
  0.59 &
  0.08 \\ \cline{2-2}
\multicolumn{1}{l|}{} &
  \multicolumn{1}{c|}{\multirow{8}{*}{CO2RR}} &
  \multicolumn{1}{c|}{$\mathrm{H_{2, Pr}}$ [$\mathrm{nmol/cm^2s}$]} &
  0.46 &
  \multicolumn{1}{c|}{87.82} &
  0.22 &
  112.67 \\
\multicolumn{1}{l|}{} &
  \multicolumn{1}{c|}{} &
  \multicolumn{1}{c|}{$\mathrm{CO_{Pr}}$} &
  0.44 &
  \multicolumn{1}{c|}{67.91} &
  0.01 &
  95.95 \\
\multicolumn{1}{l|}{} &
  \multicolumn{1}{c|}{} &
  \multicolumn{1}{c|}{$\mathrm{Liquid_{Pr}}$} &
  0.43 &
  \multicolumn{1}{c|}{55.84} &
  0.18 &
  74.11 \\
\multicolumn{1}{l|}{} &
  \multicolumn{1}{c|}{} &
  \multicolumn{1}{c|}{$\mathrm{C_2H_{4,Pr}}$} &
  0.08 &
  \multicolumn{1}{c|}{3.53} &
  0.01 &
  3.83 \\
\multicolumn{1}{l|}{} &
  \multicolumn{1}{c|}{} &
  \multicolumn{1}{c|}{$\mathrm{H_{2,FE}}$ [\%]} &
  0.48 &
  \multicolumn{1}{c|}{11.99} &
  0.17 &
  16.22 \\
\multicolumn{1}{l|}{} &
  \multicolumn{1}{c|}{} &
  \multicolumn{1}{c|}{$\mathrm{CO_{FE}}$} &
  0.55 &
  \multicolumn{1}{c|}{10.26} &
  0.10 &
  14.96 \\
\multicolumn{1}{l|}{} &
  \multicolumn{1}{c|}{} &
  \multicolumn{1}{c|}{$\mathrm{Liquid_{FE}}$} &
  0.41 &
  \multicolumn{1}{c|}{7.30} &
  0.22 &
  9.06 \\
\multicolumn{1}{l|}{} &
  \multicolumn{1}{c|}{} &
  \multicolumn{1}{c|}{$\mathrm{C_2H_{4,FE}}$} &
  0.43 &
  \multicolumn{1}{c|}{0.36} &
  0.29 &
  0.43 \\ \hline
\multicolumn{1}{l|}{\multirow{9}{*}{Spark Ablation}} &
  \multicolumn{1}{c|}{HER} &
  \multicolumn{1}{c|}{Voltage [V vs SHE]} &
  0.50 &
  \multicolumn{1}{c|}{0.08} &
  0.34 &
  0.09 \\ \cline{2-2}
\multicolumn{1}{l|}{} &
  \multicolumn{1}{c|}{\multirow{8}{*}{CO2RR}} &
  \multicolumn{1}{c|}{$\mathrm{H_{2, Pr}}$ [$\mathrm{nmol/cm^2s}$]} &
  0.45 &
  \multicolumn{1}{c|}{96.86} &
  0.26 &
  124.30 \\
\multicolumn{1}{l|}{} &
  \multicolumn{1}{c|}{} &
  \multicolumn{1}{c|}{$\mathrm{CO_{Pr}}$} &
  0.53 &
  \multicolumn{1}{c|}{60.99} &
  0.01 &
  94.53 \\
\multicolumn{1}{l|}{} &
  \multicolumn{1}{c|}{} &
  \multicolumn{1}{c|}{$\mathrm{Liquid_{Pr}}$} &
  0.45 &
  \multicolumn{1}{c|}{51.58} &
  0.03 &
  77.44 \\
\multicolumn{1}{l|}{} &
  \multicolumn{1}{c|}{} &
  \multicolumn{1}{c|}{$\mathrm{C_2H_{4,Pr}}$} &
  0.00 &
  \multicolumn{1}{c|}{3.52} &
  0.00 &
  3.60 \\
\multicolumn{1}{l|}{} &
  \multicolumn{1}{c|}{} &
  \multicolumn{1}{c|}{$\mathrm{H_{2,FE}}$ [\%]} &
  0.44 &
  \multicolumn{1}{c|}{11.31} &
  0.11 &
  15.77 \\
\multicolumn{1}{l|}{} &
  \multicolumn{1}{c|}{} &
  \multicolumn{1}{c|}{$\mathrm{CO_{FE}}$} &
  0.57 &
  \multicolumn{1}{c|}{9.81} &
  0.17 &
  13.99 \\
\multicolumn{1}{l|}{} &
  \multicolumn{1}{c|}{} &
  \multicolumn{1}{c|}{$\mathrm{Liquid_{FE}}$} &
  0.49 &
  \multicolumn{1}{c|}{5.32} &
  0.08 &
  7.82 \\
\multicolumn{1}{l|}{} &
  \multicolumn{1}{c|}{} &
  \multicolumn{1}{c|}{$\mathrm{C_2H_{4,FE}}$} &
  0.01 &
  \multicolumn{1}{c|}{0.48} &
  0.00 &
  0.55 \\ \hline
\multicolumn{1}{l|}{\multirow{9}{*}{Chemical Reduction}} &
  \multicolumn{1}{c|}{HER} &
  \multicolumn{1}{c|}{Voltage [V vs SHE]} &
  0.66 &
  \multicolumn{1}{c|}{0.07} &
  0.65 &
  0.07 \\ \cline{2-2}
\multicolumn{1}{l|}{} &
  \multicolumn{1}{c|}{\multirow{8}{*}{CO2RR}} &
  \multicolumn{1}{c|}{$\mathrm{H_{2, Pr}}$ [$\mathrm{nmol/cm^2s}$]} &
  0.30 &
  \multicolumn{1}{c|}{70.99} &
  0.12 &
  86.69 \\
\multicolumn{1}{l|}{} &
  \multicolumn{1}{c|}{} &
  \multicolumn{1}{c|}{$\mathrm{CO_{Pr}}$} &
  0.40 &
  \multicolumn{1}{c|}{69.71} &
  0.00 &
  103.27 \\
\multicolumn{1}{l|}{} &
  \multicolumn{1}{c|}{} &
  \multicolumn{1}{c|}{$\mathrm{Liquid_{Pr}}$} &
  0.33 &
  \multicolumn{1}{c|}{67.00} &
  0.27 &
  73.87 \\
\multicolumn{1}{l|}{} &
  \multicolumn{1}{c|}{} &
  \multicolumn{1}{c|}{$\mathrm{C_2H_{4,Pr}}$} &
  0.42 &
  \multicolumn{1}{c|}{3.58} &
  0.15 &
  4.02 \\
\multicolumn{1}{l|}{} &
  \multicolumn{1}{c|}{} &
  \multicolumn{1}{c|}{$\mathrm{H_{2,FE}}$ [\%]} &
  0.40 &
  \multicolumn{1}{c|}{14.63} &
  0.07 &
  19.88 \\
\multicolumn{1}{l|}{} &
  \multicolumn{1}{c|}{} &
  \multicolumn{1}{c|}{$\mathrm{CO_{FE}}$} &
  0.48 &
  \multicolumn{1}{c|}{11.17} &
  0.02 &
  16.34 \\
\multicolumn{1}{l|}{} &
  \multicolumn{1}{c|}{} &
  \multicolumn{1}{c|}{$\mathrm{Liquid_{FE}}$} &
  0.48 &
  \multicolumn{1}{c|}{8.09} &
  0.41 &
  9.37 \\
\multicolumn{1}{l|}{} &
  \multicolumn{1}{c|}{} &
  \multicolumn{1}{c|}{$\mathrm{C_2H_{4,FE}}$} &
  0.89 &
  \multicolumn{1}{c|}{0.20} &
  0.81 &
  0.27 \\ \hline
\end{tabular}%
}
\caption{Full predictive results for \cdrr~and \her~across the different synthesis techniques. A linear model using only mean adsorption energy features was used for \her~and a random forest model using Boltzmann weighted adsorption energy and Matminer features for \cdrr, representing the best results explored in this work. $\mathrm{CH_4}$ results were excluded because very little product was produced to make reliable predictive assessments.}
\label{tab:full-results}
\end{table}

\subsubsection{\her~results} \label{sec:SI_herresults}
\begin{figure}[ht!]
    \centering
    \includegraphics[width=0.65\linewidth]{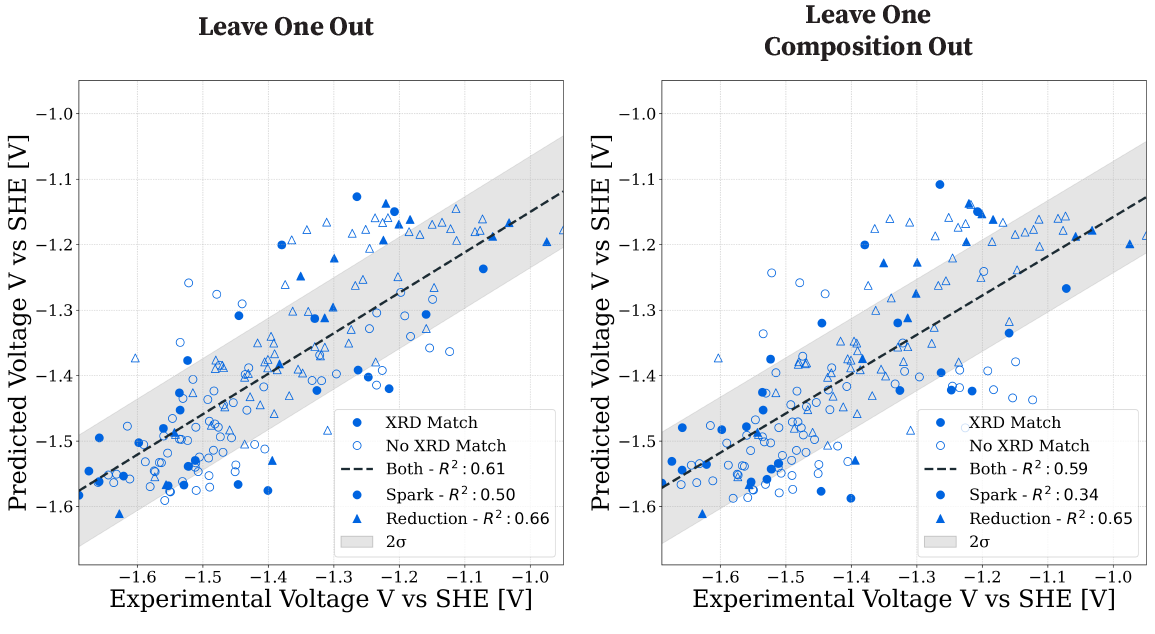}
    \caption{\her~predicted voltage for the two different cross-validation strategies. A linear model was used with features coming from mean adsorption energies.}
    \label{fig:her}
\end{figure}

\her~results evaluated against \gls{LOO} and \gls{LOCO} for a linear model using mean adsorption energy features are presented in Figure \ref{fig:her}.

\subsubsection{Ethylene results}
\begin{figure}[h!]
    \centering
    \includegraphics[width=0.65\linewidth]{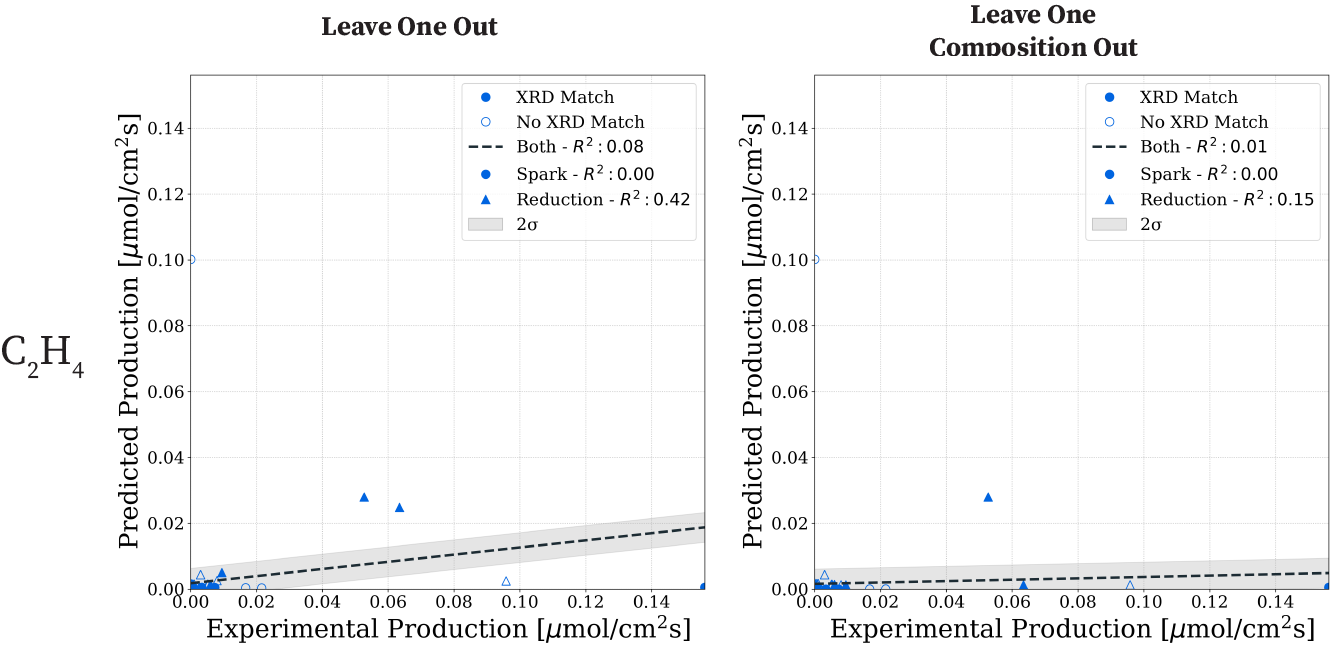}
    \caption{\cdrr~predicted production rates for \textbf{$\mathrm{C_2H_4}$} under different cross-validation strategies. A random forest regression model was used with features coming from Boltzmann weighted adsorption energies and elemental Matminer features.}
    \label{fig:c2h4}
\end{figure}

\cdrr~results evaluated on $\mathrm{C_2H_4}$ are provided in Figure \ref{fig:c2h4}. The limited availability of $\mathrm{C_{2+}}$ producing samples makes the fitting not particularly meaningful. Similar results are also observed with $\mathrm{CH_4}$.

\subsubsection{Linear model}
\begin{figure}[h!]
    \centering
    \includegraphics[width=0.88\linewidth]{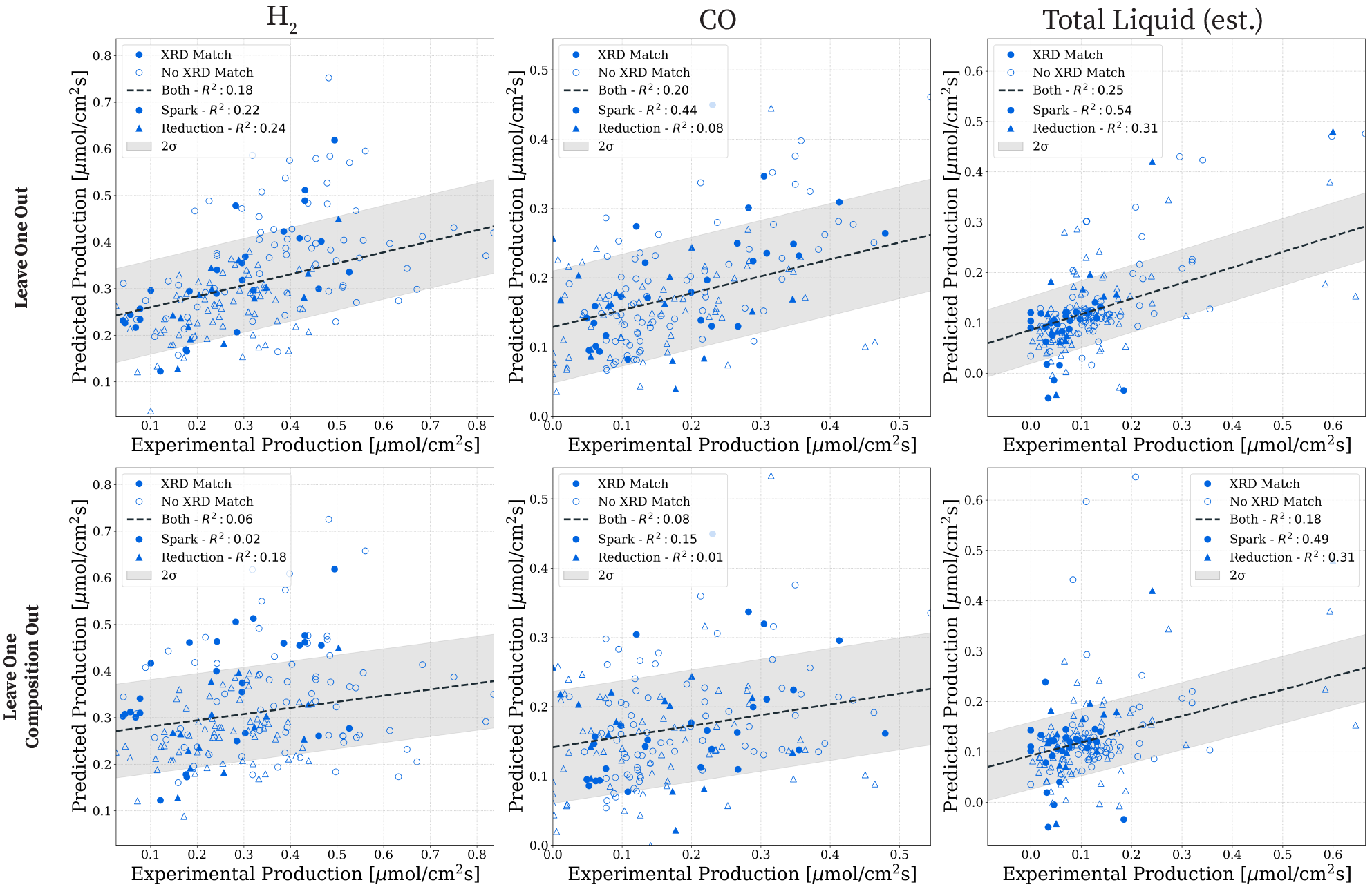}
    \caption{\gls{CO2RR} predicted production rates across several products - $\mathrm{H_2}$, CO, and Total Liquid for the two different cross-validation strategies. A \textbf{linear model} was used with features coming from Boltzmann weighted adsorption energies and elemental Matminer features.}
    \label{fig:linear}
\end{figure}

The main paper presented \cdrr~results trained on a random forest model. We also present results using a linear model in Figure \ref{fig:linear}. The results are considerably worse than random forest. 

\subsubsection{Matminer-only features}
\begin{figure}[h!]
    \centering
    \includegraphics[width=0.88\linewidth]{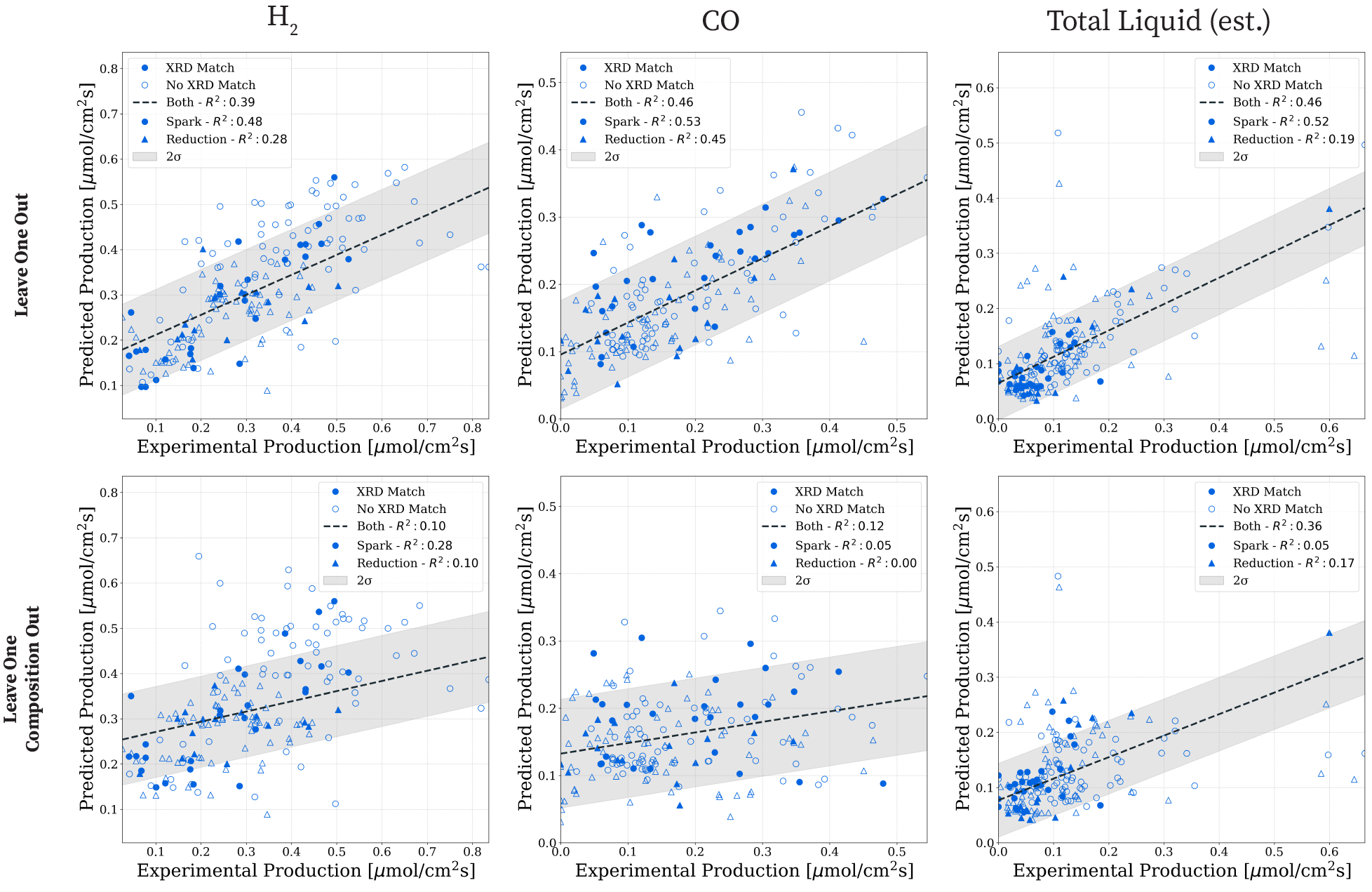}
    \caption{\gls{CO2RR} predicted production rates across several products - $\mathrm{H_2}$, CO, and Total Liquid for the two different cross-validation strategies. A random forest regression model was used with \textbf{only elemental Matminer features.}}
    \label{fig:matminer}
\end{figure}

The main paper presented \cdrr~results trained using both Boltzmann-weighted adsorption energies and elemental Matminer features. Figure \ref{fig:matminer} shows the performance using only Matminer features. The results using only elemental features are very comparable to those also considering adsorption energies. 

\subsubsection{Matched-only features}
\begin{figure}[h!]
    \centering
    \includegraphics[width=0.88\linewidth]{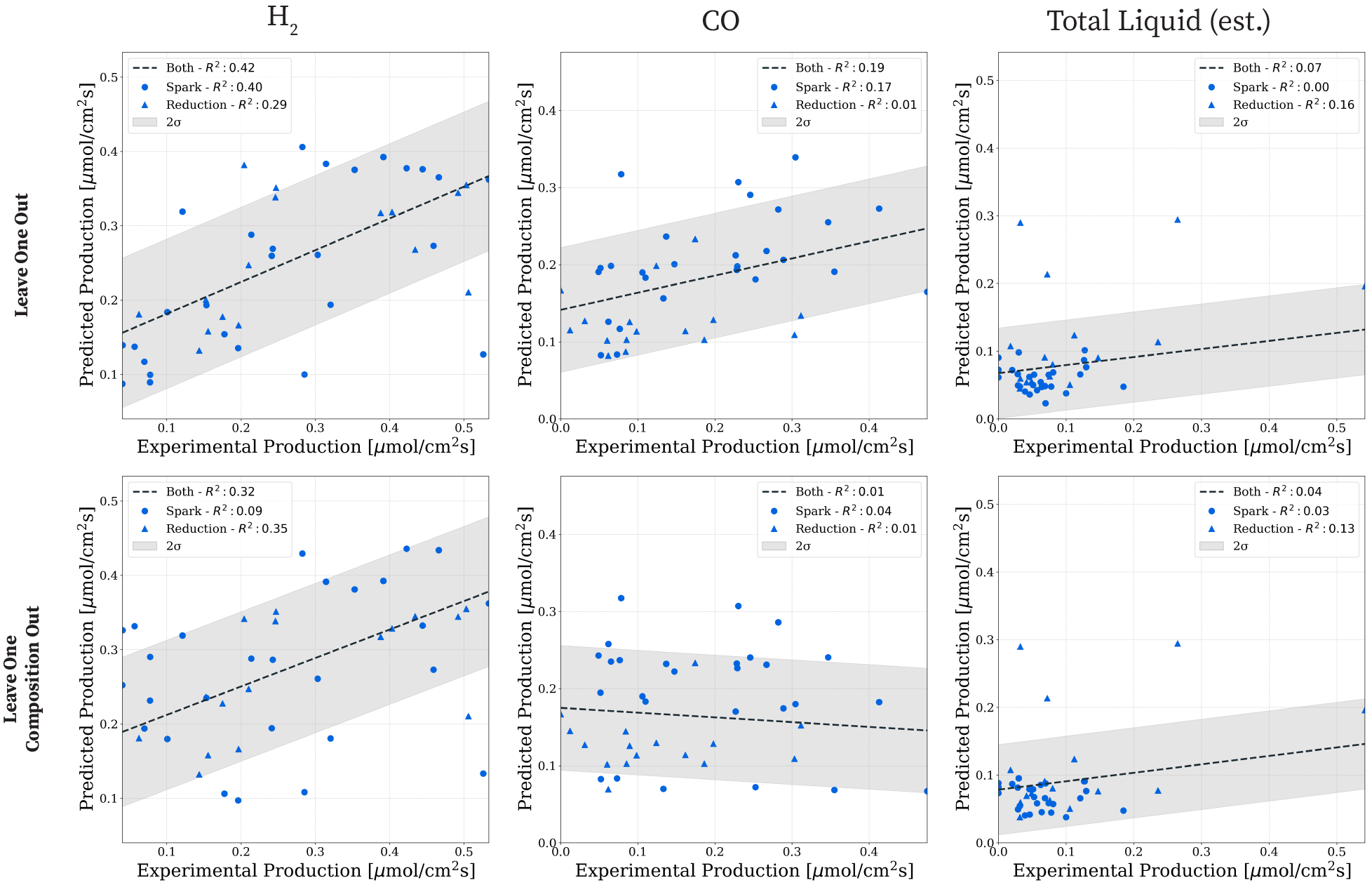}
    \caption{\gls{CO2RR} predicted production rates across several products - $\mathrm{H_2}$, CO, and Total Liquid for the two different cross-validation strategies. Only \textbf{XRD matched} samples were considered in this analysis.}
    \label{fig:match}
\end{figure}

The main paper presented \cdrr~results fitted on both \gls{XRD} matched and unmatched data. The results of fitting a model on only the \gls{XRD} matched data are shown in Figure \ref{fig:match}. Fitting on both matched and unmatched data tends to help our predictive models and they are more performant than only fitting on \gls{XRD} matched data. This is likely a result of both categories being still far from the idealized computational model.

\subsubsection{Faradaic efficiency results}
\begin{figure}[h!]
    \centering
    \includegraphics[width=0.65\linewidth]{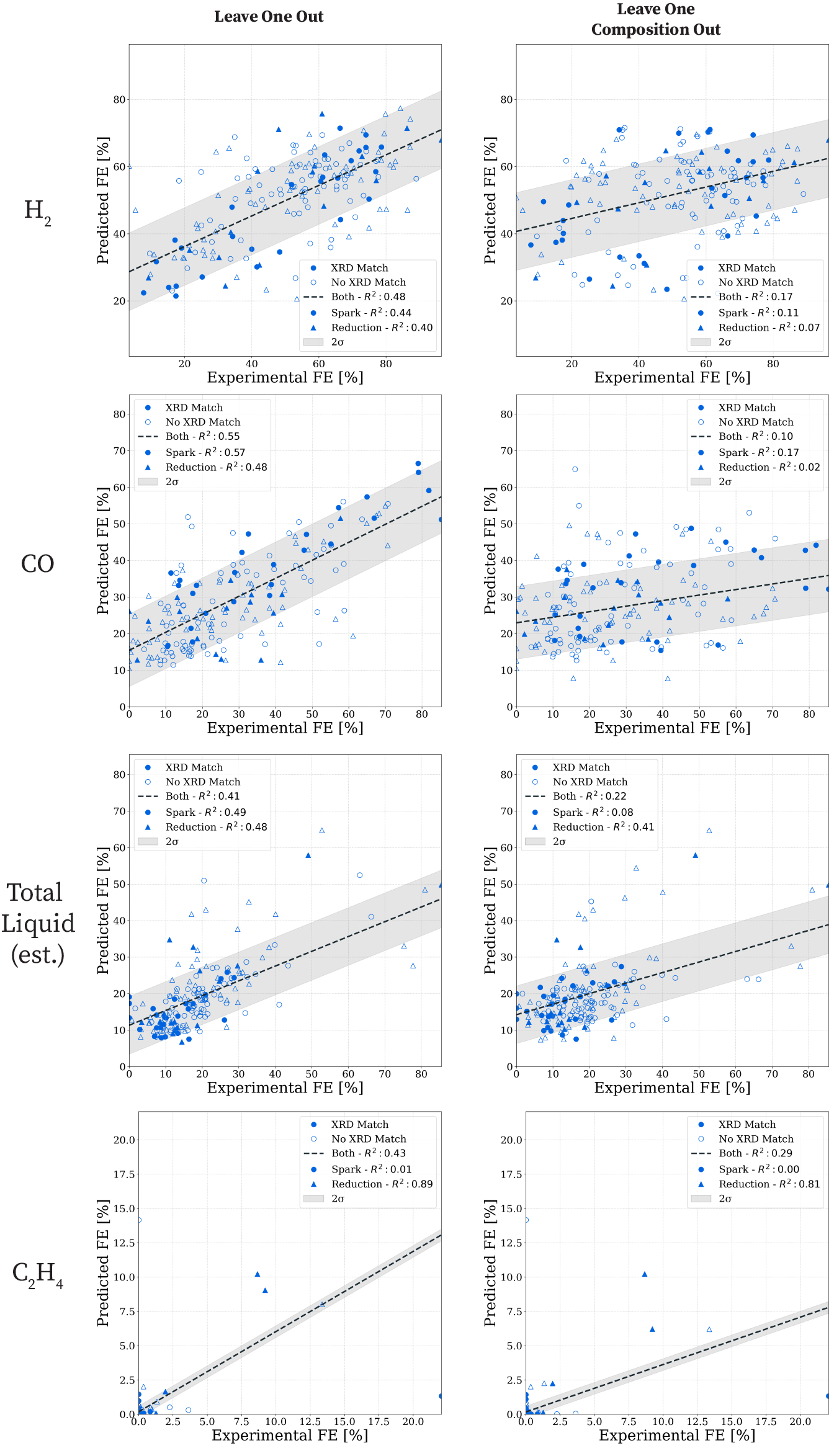}
    \caption{\gls{CO2RR} predicted \textbf{Faradaic efficiencies} across several products - $\mathrm{H_2}$, CO, Total Liquid, and $\mathrm{C_2H_4}$ for the two different cross-validation strategies. A random forest model was used with features coming from Boltzmann weighted adsorption energies and elemental Matminer features.}
    \label{fig:fe}
\end{figure}

Instead of fitting on production rates, we can also try fitting on the Faradaic efficiencies of the different products . Results for the ~\cdrr~products are shown in Figure \ref{fig:fe}. Overall, results on \gls{LOO} are on par and some instances better (CO - 0.55 vs. 0.44) than those trained for production rate. However, the same is not true for \gls{LOCO}, where results are mixed. Both sets of results still suggest a very poor correlation.

\clearpage

\subsection{HER Candidates} \label{sec:HER_candidates}
\tiny{
\begin{longtable}{llrllr}
\toprule
           material id &                  composition &  predicted voltage &           material id &                  composition &    predicted voltage\\
            & & V v. SHE & & & V v. SHE \\
\midrule
      oqmd-1128070 &      Mo-0.25-Nb-0.25-W-0.5 &                   0.20 & dcgat-agm002185003 &                              Se-1.0 & -1.17 \\
        mp-1187206 &             Ta-0.75-W-0.25 &                  -0.80 &         mp-1071078 &                   Ni-0.333-Se-0.667 & -1.17 \\
       oqmd-447729 &       Mo-0.5-V-0.25-W-0.25 &                  -0.91 & dcgat-agm002183076 &                        Cd-0.5-S-0.5 & -1.17 \\
          mp-20456 &          Co-0.429-Se-0.571 &                  -0.92 &          mp-973053 &                      H-0.75-Sc-0.25 & -1.17 \\
       oqmd-472683 &     Ga-0.25-Mn-0.5-Zn-0.25 &                  -0.95 & dcgat-agm003184034 &              Pb-0.25-Se-0.5-Tl-0.25 & -1.17 \\
dcgat-agm002153716 &           Ge-0.111-H-0.889 &                  -0.96 & dcgat-agm003062322 &              Cu-0.25-Se-0.5-Zn-0.25 & -1.17 \\
       oqmd-520732 &     Ga-0.5-Mn-0.25-Tc-0.25 &                  -0.98 & dcgat-agm003153632 &              Hg-0.25-Se-0.5-Zn-0.25 & -1.17 \\
       oqmd-393713 &      Co-0.25-Mo-0.25-V-0.5 &                  -0.98 &          mp-580226 &                       Cu-0.5-Se-0.5 & -1.17 \\
        mp-1030108 &   Mo-0.083-Te-0.667-W-0.25 &                  -1.00 &         mp-1063938 &                       As-0.5-Mn-0.5 & -1.17 \\
          mp-11501 &            Mn-0.25-Ni-0.75 &                  -1.02 & dcgat-agm003144793 &                       Bi-0.4-Se-0.6 & -1.17 \\
         mp-600124 & Mn-0.333-Ni-0.333-Sb-0.333 &                  -1.03 &         mp-1103177 &                   Fe-0.333-Se-0.667 & -1.18 \\
      oqmd-1222216 &              Cu-0.5-Se-0.5 &                  -1.03 & dcgat-agm003738920 &          Sb-0.167-Se-0.667-Sn-0.167 & -1.18 \\
dcgat-agm002183072 &               Cd-0.5-S-0.5 &                  -1.03 &       oqmd-1555855 &              Hg-0.5-Se-0.25-Te-0.25 & -1.18 \\
dcgat-agm002185503 & Hg-0.143-Se-0.571-Tl-0.286 &                  -1.04 &           mp-22745 &                   Co-0.529-Se-0.471 & -1.18 \\
dcgat-agm002185494 & Cd-0.143-Se-0.571-Tl-0.286 &                  -1.05 & dcgat-agm003204187 &                   Se-0.667-Tl-0.333 & -1.18 \\
dcgat-agm003062314 &     Cu-0.25-Se-0.5-Zn-0.25 &                  -1.05 & dcgat-agm003062313 &              Cu-0.25-Se-0.5-Zn-0.25 & -1.18 \\
dcgat-agm002193870 & Se-0.571-Tl-0.286-Zn-0.143 &                  -1.06 & dcgat-agm002370150 &                   Co-0.167-Se-0.833 & -1.18 \\
dcgat-agm002184996 &                     Se-1.0 &                  -1.06 &            mp-1821 &                    Se-0.667-W-0.333 & -1.18 \\
dcgat-agm002019659 &                     Se-1.0 &                  -1.06 & dcgat-agm003659759 &                Sb-0.1-Se-0.5-Tl-0.4 & -1.18 \\
dcgat-agm003201183 &              Se-0.5-Tl-0.5 &                  -1.08 &         mp-1226897 &                Cd-0.4-Hg-0.1-Se-0.5 & -1.18 \\
dcgat-agm002019668 &                     Se-1.0 &                  -1.08 & dcgat-agm003198330 &                   Ni-0.167-Se-0.833 & -1.18 \\
dcgat-agm003224862 &               Ge-0.2-H-0.8 &                  -1.08 & dcgat-agm002071817 &                    H-0.667-Sb-0.333 & -1.18 \\
      oqmd-1441866 &              Cu-0.5-Se-0.5 &                  -1.08 &         mp-1226039 &          Co-0.167-Ni-0.167-Te-0.667 & -1.18 \\
dcgat-agm002069646 &           As-0.333-H-0.667 &                  -1.08 &         mp-1226036 &          As-0.333-Co-0.333-Se-0.333 & -1.18 \\
dcgat-agm003198763 &          Se-0.667-Tl-0.333 &                  -1.08 & dcgat-agm003218501 &                   Se-0.375-Tl-0.625 & -1.18 \\
            mp-820 &              Hg-0.5-Se-0.5 &                  -1.09 & dcgat-agm003313875 &          Se-0.556-Te-0.222-Tl-0.222 & -1.18 \\
dcgat-agm003143322 &     Cd-0.25-Hg-0.25-Se-0.5 &                  -1.10 & dcgat-agm003143331 &              Cd-0.25-Hg-0.25-Se-0.5 & -1.18 \\
dcgat-agm002185495 & Cd-0.143-Se-0.571-Tl-0.286 &                  -1.10 & dcgat-agm002018208 &                   Se-0.667-Te-0.333 & -1.18 \\
dcgat-agm002183071 &               Cd-0.5-S-0.5 &                  -1.10 & dcgat-agm003765687 &           Co-0.125-Cu-0.125-Se-0.75 & -1.18 \\
dcgat-agm003143326 &     Cd-0.25-Hg-0.25-Se-0.5 &                  -1.10 & dcgat-agm002239472 &                       In-0.4-Se-0.6 & -1.19 \\
      oqmd-1754148 &              Cu-0.5-Se-0.5 &                  -1.10 & dcgat-agm001623308 &          Cu-0.2-H-0.2-Mn-0.2-Ni-0.4 & -1.19 \\
         mp-680646 &              Ni-0.5-Sn-0.5 &                  -1.10 & dcgat-agm002494639 &                 Co-0.2-H-0.6-Nb-0.2 & -1.19 \\
         mp-673255 &          Cu-0.529-Se-0.471 &                  -1.10 & dcgat-agm003596580 &          Se-0.444-Te-0.111-Tl-0.444 & -1.19 \\
         mp-866134 &             Fe-0.75-V-0.25 &                  -1.11 &         mp-1223932 &              Hg-0.5-Se-0.25-Te-0.25 & -1.19 \\
      oqmd-1105090 &              Se-0.5-Tl-0.5 &                  -1.11 & dcgat-agm002017918 &                   Bi-0.333-Se-0.667 & -1.19 \\
        mp-1025649 &  Mo-0.222-Te-0.667-W-0.111 &                  -1.11 & dcgat-agm003625178 &              Cd-0.25-Hg-0.25-Se-0.5 & -1.19 \\
       oqmd-297717 &            Co-0.75-Ni-0.25 &                  -1.11 & dcgat-agm002018271 &                     In-0.25-Se-0.75 & -1.19 \\
dcgat-agm003292803 &          Se-0.455-Tl-0.545 &                  -1.11 &         oqmd-20460 &          As-0.333-Mn-0.333-Ni-0.333 & -1.19 \\
dcgat-agm002268684 &          Cu-0.333-Se-0.667 &                  -1.12 &          oqmd-7853 &                Cu-0.4-Se-0.4-Tl-0.2 & -1.19 \\
dcgat-agm002019672 &                     Se-1.0 &                  -1.12 &         mp-1225994 &          Co-0.333-Se-0.333-Te-0.333 & -1.19 \\
dcgat-agm002193869 & Se-0.571-Tl-0.286-Zn-0.143 &                  -1.12 &       oqmd-1480903 &          As-0.333-Bi-0.333-Sb-0.333 & -1.19 \\
dcgat-agm002888614 &     Cd-0.25-Hg-0.25-Se-0.5 &                  -1.12 &       oqmd-1739776 &                   Mo-0.333-Se-0.667 & -1.19 \\
dcgat-agm003061770 &     Cu-0.25-Hg-0.25-Se-0.5 &                  -1.12 & dcgat-agm003193011 &              Se-0.5-Te-0.25-Tl-0.25 & -1.19 \\
dcgat-agm002183073 &               Cd-0.5-S-0.5 &                  -1.12 & dcgat-agm002295867 &                   Ni-0.333-Se-0.667 & -1.19 \\
       oqmd-322981 &            Mn-0.25-Ni-0.75 &                  -1.12 &             mp-672 &                        Cd-0.5-S-0.5 & -1.19 \\
dcgat-agm003556772 &   In-0.125-Se-0.5-Tl-0.375 &                  -1.12 &         mp-1027580 &           Mo-0.333-S-0.333-Se-0.333 & -1.19 \\
      oqmd-1368161 &   Cu-0.333-Se-0.5-Tl-0.167 &                  -1.12 & dcgat-agm003676550 &            Pb-0.417-Se-0.5-Tl-0.083 & -1.19 \\
dcgat-agm003212411 &          Se-0.667-Tl-0.333 &                  -1.12 & dcgat-agm003556315 &             Cd-0.5-S-0.375-Se-0.125 & -1.19 \\
        mp-1026351 &  Mo-0.111-Te-0.667-W-0.222 &                  -1.12 & dcgat-agm003192331 &              Hg-0.25-Se-0.5-Te-0.25 & -1.19 \\
        mp-1226731 &       Cd-0.1-Hg-0.4-Se-0.5 &                  -1.12 &         mp-1212695 &          Ga-0.133-Hg-0.333-Se-0.533 & -1.19 \\
           mp-1836 &              Se-0.5-Tl-0.5 &                  -1.13 &         mp-1025906 &           Mo-0.333-S-0.222-Se-0.444 & -1.19 \\
        mp-1025819 &  Mo-0.333-S-0.222-Se-0.444 &                  -1.13 & dcgat-agm002018209 &                   Se-0.667-Te-0.333 & -1.19 \\
dcgat-agm003556575 &   Cu-0.375-Ni-0.125-Se-0.5 &                  -1.13 &        oqmd-717228 &              Cu-0.5-Ni-0.25-Zn-0.25 & -1.20 \\
          mp-14090 &     Cu-0.25-Se-0.5-Tl-0.25 &                  -1.13 &         mp-1201566 &            As-0.125-Se-0.5-Tl-0.375 & -1.20 \\
           mp-2280 &          Cu-0.333-Se-0.667 &                  -1.13 &           mp-20901 &                   Ni-0.333-Se-0.667 & -1.20 \\
           mp-2000 &          Cu-0.333-Se-0.667 &                  -1.13 & dcgat-agm003062327 &              Cu-0.25-Se-0.5-Zn-0.25 & -1.20 \\
dcgat-agm002069645 &           As-0.333-H-0.667 &                  -1.13 &         mp-1065751 &              Co-0.5-Fe-0.25-Si-0.25 & -1.20 \\
dcgat-agm003203021 &                     Se-1.0 &                  -1.13 &        oqmd-552445 &              Sn-0.25-Ta-0.25-Tc-0.5 & -1.20 \\
dcgat-agm003293190 &          Se-0.417-Tl-0.583 &                  -1.13 &       oqmd-1440299 &                   Ni-0.333-Se-0.667 & -1.20 \\
dcgat-agm002183074 &               Cd-0.5-S-0.5 &                  -1.14 & dcgat-agm002152745 &          Bi-0.333-Ge-0.083-Se-0.583 & -1.20 \\
dcgat-agm003267594 &           H-0.889-Si-0.111 &                  -1.14 &            mp-4492 &              Co-0.5-Mn-0.25-Si-0.25 & -1.20 \\
dcgat-agm001215469 &     Ge-0.25-Mn-0.25-Ni-0.5 &                  -1.14 &         mp-1212307 &                       In-0.7-Ni-0.3 & -1.20 \\
       oqmd-308096 &             Mo-0.25-W-0.75 &                  -1.14 & dcgat-agm002167041 &                   As-0.333-Ni-0.667 & -1.20 \\
dcgat-agm003676549 &   Sb-0.083-Se-0.5-Tl-0.417 &                  -1.14 & dcgat-agm002151080 &                       Ga-0.4-Se-0.6 & -1.20 \\
      oqmd-1473251 &              Hg-0.5-Se-0.5 &                  -1.14 & dcgat-agm002153596 &           Fe-0.111-H-0.667-Zn-0.222 & -1.20 \\
dcgat-agm003183055 &     Mn-0.25-Ni-0.5-Zn-0.25 &                  -1.14 & dcgat-agm003153634 &              Hg-0.25-Se-0.5-Zn-0.25 & -1.20 \\
dcgat-agm002018195 &          Se-0.667-Te-0.333 &                  -1.14 &            mp-9814 &            Cu-0.375-Sb-0.125-Se-0.5 & -1.20 \\
dcgat-agm003659717 &       Se-0.5-Te-0.1-Tl-0.4 &                  -1.14 &       oqmd-1741060 &            Bi-0.125-Cu-0.375-Se-0.5 & -1.20 \\
dcgat-agm003289670 &          Se-0.429-Tl-0.571 &                  -1.15 &            mp-5396 &              Co-0.5-Mn-0.25-Sb-0.25 & -1.20 \\
dcgat-agm002018216 &          Se-0.667-Te-0.333 &                  -1.15 &            mp-2090 &                       Co-0.5-Fe-0.5 & -1.20 \\
      oqmd-1369425 &           Mn-0.333-V-0.667 &                  -1.15 &           mp-10074 &                   Ge-0.333-Se-0.667 & -1.20 \\
        mp-1080603 &          Co-0.125-Fe-0.875 &                  -1.15 & dcgat-agm002018272 &                     In-0.25-Se-0.75 & -1.21 \\
dcgat-agm002184995 &                     Se-1.0 &                  -1.15 & dcgat-agm003414366 &            Se-0.5-Te-0.167-Tl-0.333 & -1.21 \\
dcgat-agm002018198 &          Se-0.667-Te-0.333 &                  -1.15 &            mp-2578 &                   Ni-0.333-Te-0.667 & -1.21 \\
      oqmd-1752850 &   Sb-0.125-Se-0.5-Tl-0.375 &                  -1.15 & dcgat-agm003615401 &              Cu-0.5-Se-0.25-Te-0.25 & -1.21 \\
dcgat-agm003153628 &     Hg-0.25-Se-0.5-Zn-0.25 &                  -1.15 & dcgat-agm003257709 & Cu-0.143-Ga-0.143-Se-0.571-Sn-0.143 & -1.21 \\
         mp-982261 &          Pb-0.333-Se-0.667 &                  -1.15 &          mp-675626 &            As-0.125-Cu-0.375-Se-0.5 & -1.21 \\
          mp-22309 &          Co-0.333-Se-0.667 &                  -1.15 &           mp-22300 &              Co-0.5-Fe-0.25-Ge-0.25 & -1.21 \\
        mp-1226729 &     Cd-0.25-Hg-0.25-Se-0.5 &                  -1.15 & dcgat-agm002264007 &            Co-0.125-Ni-0.375-Se-0.5 & -1.21 \\
         mp-568380 &          Se-0.375-Tl-0.625 &                  -1.15 &         mp-1226990 &          Cd-0.273-In-0.182-Se-0.545 & -1.21 \\
dcgat-agm003354875 & Se-0.615-Te-0.231-Tl-0.154 &                  -1.16 & dcgat-agm002018213 &                   Se-0.667-Te-0.333 & -1.21 \\
dcgat-agm003676192 &   Se-0.5-Sn-0.083-Tl-0.417 &                  -1.16 &       oqmd-1735308 &            Sb-0.125-Se-0.5-Tl-0.375 & -1.21 \\
       oqmd-551623 &     Ga-0.25-Ta-0.25-Tc-0.5 &                  -1.16 &       oqmd-1473593 &                   Se-0.667-Te-0.333 & -1.21 \\
dcgat-agm003455149 & Hg-0.143-Se-0.571-Tl-0.286 &                  -1.16 &       oqmd-1339137 &                   Ga-0.857-Mn-0.143 & -1.21 \\
dcgat-agm002144077 &          Co-0.273-Se-0.727 &                  -1.16 &       oqmd-1473632 &                    Te-0.667-W-0.333 & -1.21 \\
dcgat-agm002345642 &          Mn-0.167-Se-0.833 &                  -1.16 &          mp-568661 &          Cd-0.143-In-0.286-Se-0.571 & -1.21 \\
        mp-1018722 &              Hg-0.5-Se-0.5 &                  -1.16 &       oqmd-1440294 &                   Se-0.667-Te-0.333 & -1.21 \\
           mp-2469 &               Cd-0.5-S-0.5 &                  -1.16 & dcgat-agm003194428 &                       As-0.4-Se-0.6 & -1.21 \\
dcgat-agm003556778 &   Bi-0.125-Se-0.5-Tl-0.375 &                  -1.16 &           mp-20862 &                   Co-0.333-Se-0.667 & -1.21 \\
dcgat-agm003429338 &   Pb-0.333-Se-0.5-Tl-0.167 &                  -1.16 &            mp-2730 &                       Hg-0.5-Te-0.5 & -1.21 \\
dcgat-agm003654045 &       Pb-0.1-Se-0.4-Tl-0.5 &                  -1.16 &       oqmd-1017516 &              Co-0.5-Fe-0.25-Ge-0.25 & -1.21 \\
         mp-601833 &          Fe-0.812-Ge-0.188 &                  -1.17 &         mp-1224765 &           Fe-0.25-Ni-0.083-Sb-0.667 & -1.21 \\
dcgat-agm002185433 &              Se-0.6-Sn-0.4 &                  -1.17 & dcgat-agm002018194 &                   Se-0.667-Te-0.333 & -1.22 \\
dcgat-agm002018273 &            In-0.25-Se-0.75 &                  -1.17 &       oqmd-1739029 &                   Mo-0.333-Se-0.667 & -1.22 \\
        mp-1026916 &  Mo-0.333-S-0.333-Se-0.333 &                  -1.17 &           mp-31406 &                Bi-0.4-Se-0.4-Te-0.2 & -1.22 \\
dcgat-agm002183075 &               Cd-0.5-S-0.5 &                  -1.17 &       oqmd-1474575 &            Sb-0.125-Se-0.5-Tl-0.375 & -1.22 \\
dcgat-agm003143324 &     Cd-0.25-Hg-0.25-Se-0.5 &                  -1.17 &         mp-1025588 &            S-0.222-Se-0.444-W-0.333 & -1.22 \\
dcgat-agm002381699 &            Pb-0.25-Se-0.75 &                  -1.17 &           mp-22784 &                     In-0.25-Ni-0.75 & -1.22 \\
dcgat-agm002146833 &          Cu-0.333-Se-0.667 &                  -1.17 &       oqmd-1443229 &            As-0.125-Cu-0.375-Se-0.5 & -1.22 \\
      oqmd-1736110 & Hg-0.143-In-0.286-Se-0.571 &                  -1.17 & dcgat-agm002017915 &                   Bi-0.333-Se-0.667 & -1.22 \\
\bottomrule
\end{longtable}
}

% \printglossary

\end{document}